\documentclass[10pt,twocolumn,superscriptaddress,amssymb, amsmath, aps, showpacs, footinbib,  prb]{revtex4-1}
\usepackage{graphicx,color}

\newcommand{\eff}{\text{eff}}

\newcommand{\AFM}{\text{AFM}}

\newcommand{\ra}{\rangle}
\newcommand{\Neel}{N\'{e}el }
\newcommand{\bea}{\begin{eqnarray}}
\newcommand{\eea}{\end{eqnarray}}
\newcommand{\kv}{\mathbf k}

\begin{document}

\title{Spiral ground state in the quasi-two-dimensional spin-$\frac12$ system Cu$_2$GeO$_4$}

\author{Alexander A. Tsirlin}
\email{altsirlin@gmail.com}
\affiliation{Max Planck Institute for Chemical Physics of Solids, N\"{o}thnitzer
Str. 40, 01187 Dresden, Germany}
\author{Ronald Zinke}
\author{Johannes Richter}
\affiliation{Institute for Theoretical Physics, University of Magdeburg, P.O. Box 4120,
39016 Magdeburg, Germany}
\author{Helge Rosner}
\email{Helge.Rosner@cpfs.mpg.de}
\affiliation{Max Planck Institute for Chemical Physics of Solids, N\"{o}thnitzer
Str. 40, 01187 Dresden, Germany}

\begin{abstract}
We apply density functional theory band structure calculations, the coupled-cluster method, and exact diagonalization to investigate the microscopic magnetic model of the spin-$\frac12$ compound Cu$_2$GeO$_4$. The model is quasi-two-dimensional, with uniform spin chains along one direction and frustrated spin chains along the other direction. The coupling along the uniform chains is antiferromagnetic, $J\simeq 130$~K. The couplings along the frustrated chains are $J_1\simeq -60$~K and $J_2\simeq 80$~K between nearest neighbors and next-nearest neighbors, respectively. The ground state of the quantum model is a spiral, with the reduced sublattice magnetization of 0.62~$\mu_B$ and the pitch angle of $84^{\circ}$, both renormalized by quantum effects. The proposed spiral ground state of Cu$_2$GeO$_4$ opens a way to magnetoelectric effects in this compound.
\end{abstract}

\maketitle

\section{Introduction}
Quantum magnetism is a field of fundamental research focused on exotic ground states and non-trivial low-temperature properties.\cite{frustrated,quantum} Nevertheless, certain effects in quantum magnets are also relevant for applications. Spin-chain compounds show ballistic regime 
of heat transport,\cite{ballistic} whereas frustrated magnets are capable of a strong 
magnetocaloric effect.\cite{zhitomirsky2004,schnack2007} Additionally, many of the frustrated magnets undergo spiral or, in general, incommensurate ordering, and reveal ferroelectricity induced by a magnetic field.\cite{mostovoy}  A frustrated spin chain with competing ferromagnetic (FM) nearest-neighbor ($J_1$) and antiferromagnetic (AFM) next-nearest-neighbor ($J_2$) couplings is the simplest spin model giving rise to spiral magnetic correlations at $J_2/J_1<-\frac14$ (Ref.~\onlinecite{furukawa2010}). This model is easily realized experimentally and has a clear structural footprint, a chain of edge-sharing CuX$_4$ plaquettes with X being oxygen,\cite{mazurenko2007,gippius2004,*masuda2005,licuvo4,*comment,*enderle2010,capogna2005,*drechsler2006} chlorine,\cite{cucl2,*banks2009} or even nitrogen.\cite{tsirlin2010} Such chains typically show FM $J_1$ due to the nearly $90^{\circ}$ Cu--X--Cu angle and AFM $J_2$ due to the Cu--X--X--Cu superexchange. Indeed, many compounds of this type undergo spiral magnetic ordering and sometimes exhibit magnetic field-induced ferroelectricity.\cite{park2007,*naito2007,*seki2010} However, the detailed microscopic understanding of these effects remains challenging, and even the electronic origin of ferroelectricity in spin-chain cuprates is vividly debated.\cite{moskvin,*moskvin-2}

Interchain couplings are an important feature of any real material. The couplings between spin chains can modify the ground state qualitatively by inducing a long-range order with finite sublattice magnetization.\cite{affleck1994,sandvik1999} In the case of frustrated spin chains, such couplings influence the behavior of doped systems,\cite{laflorencie2003} and play a decisive role for the stability of exotic phases in high magnetic fields.\cite{ueda2009,*zhitomirsky,nishimoto2010} Regarding magnetoelectric effects, the interchain couplings naturally determine their temperature scale by adjusting the magnetic ordering temperature. 

Theoretical studies of coupled frustrated spin chains remain a challenge owing to the two-dimensional (2D) and frustrated nature of the problem. Therefore, experimental benchmarks are especially important. The available frustrated-spin-chain compounds show relatively weak interchain couplings,\cite{nishimoto2010} while the relevance of the opposite regime with strongly coupled frustrated spin chains remains unclear. A common and a somewhat naive picture suggests that leading exchange couplings should run along the structural chains owing to shorter Cu--Cu distances.\cite{zinke2009} 

\begin{figure}
\includegraphics{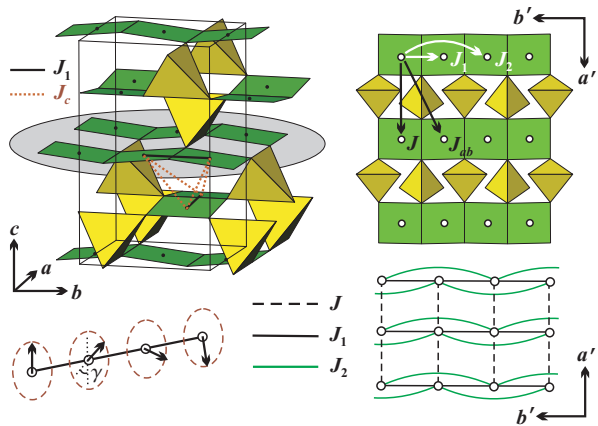}
\caption{\label{fig:structure}
(Color online) Top panel: crystal structure of Cu$_2$GeO$_4$ (left) and a single magnetic layer in the $ab$ plane (right). Bottom panel: a sketch of the spin spiral with the 
pitch angle $\gamma$ (left), and the magnetic model of $J_1-J_2$ frustrated spin chains coupled by $J$ (right). Circles and dots denote the positions of the Cu atoms. Lines in the top left panel show the anisotropic pyrochlore lattice considered in 
Ref.~\onlinecite{yamada2000}.
}
\end{figure}

In the following, we present a microscopic magnetic model of Cu$_2$GeO$_4$. This compound is a unique example of a 2D system of strongly coupled frustrated spin chains. The coupling $J$ between the frustrated chains is so strong that the system can be equally viewed as uniform spin chains along $J$ with the frustrated interchain couplings $J_1$ and $J_2$ (see Fig.~\ref{fig:structure}).  Both descriptions relate to certain features of the magnetic behavior: while the uniform-chain model fits the magnetic susceptibility of Cu$_2$GeO$_4$ down to $T/J\simeq 0.5$, the ground state of the 2D model is a spiral, which is typical for the frustrated $J_1-J_2$ spin chains.

The crystal structure of Cu$_2$GeO$_4$ belongs to the spinel type.\cite{cu2geo4-str} Magnetic properties were studied in a relation to the spin-Peierls compound CuGeO$_3$. The low-dimensional magnetic behavior of Cu$_2$GeO$_4$ resembles CuGeO$_3$ indeed. However, no signatures of the structural distortion or spin gap were found down to 10~K, and the long-range magnetic ordering at $T_N=33.1$~K is observed instead.\cite{yamada2000}

Yamada \textit{et al}.\cite{yamada2000} analyzed Cu$_2$GeO$_4$ using the anisotropic pyrochlore lattice model with two inequivalent exchange couplings that are $J_1$ and $J_c$ in our notation (upper left panel of Fig.~\ref{fig:structure}). This model arises from a straight-forward and naive geometrical consideration of the spinel structure, with inequivalent couplings driven by the tetragonal distortion of the parent cubic system. At $J_c/J_1\ll 1$, the anisotropic pyrochlore lattice splits into chains. According to Ref.~\onlinecite{yamada2000}, Cu$_2$GeO$_4$ is close to this limit, with $J_1=135$~K and $J_c/J_1=0.16$. Starykh \textit{et al}.\cite{starykh2005} studied the 2D analog of the model theoretically, and proposed a quantum-disordered valence-bond-solid ground state.

\section{Band structure}
\label{sec:band}
As a derivative of the spinel structure, Cu$_2$GeO$_4$ might be thought of as a three-dimensional network of CuO$_6$ octahedra. However, this description ignores essential features of the electronic structure. In oxide compounds, Cu$^{+2}$ tends to adopt a four-fold coordination (CuO$_4$ plaquette) having dramatic influence on the orbital ground state and magnetic properties. Such plaquettes can be recognized in Cu$_2$GeO$_4$, and lead to a peculiar superexchange scenario. Four short bonds to oxygen (1.95~\r A) form the CuO$_4$ plaquettes in the $ab$ plane, whereas the two remaining Cu--O bonds are much longer (2.50~\r A). Edge-sharing CuO$_4$ plaquettes comprise structural chains that run along $a$ or $b$, with parallel chains forming layers in the $ab$ plane (upper right panel of Fig.~\ref{fig:structure}). Equivalent layers with differently directed structural chains alternate along the $c$ axis. In the following, we denote the direction of the structural chains as $b'$ and the perpendicular direction as $a'$, to distinguish those from the crystallographic $a$ and $b$ axes. GeO$_4$ tetrahedra connect the chains into a three-dimensional (3D) framework (Fig.~\ref{fig:structure}).

To evaluate individual exchange couplings, we perform scalar-relativistic density functional theory (DFT) band structure calculations using the \texttt{FPLO9.00-33} code.\cite{fplo} We apply the local density approximation (LDA) with the exchange-correlation potential by Perdew and Wang,\cite{pw92} and use a well-converged $k$ mesh comprising 3350 points in the symmetry-irreducible part of the first Brillouin zone. With LDA calculations, we are able to identify relevant states, and to evaluate hopping parameters 
$t_i$ via a fit with an effective one-orbital tight-binding (TB) model. The hopping parameters are introduced into a Hubbard model with the effective on-site Coulomb repulsion potential $U_{\eff}=4.5$~eV.\cite{dioptase,cu2v2o7,janson2009} In the case of low-lying excitations, the Hubbard model is further reduced to a Heisenberg model under the conditions of half-filling and strong correlations ($t_i\ll U_{\eff}$). Then, the AFM parts of the exchange integrals are evaluated as $J_i^{\AFM}=4t_i^2/U_{\eff}$. 

An alternative way to evaluate the exchange couplings is to treat the strong correlations within DFT, via the mean-field-like LSDA+$U$ approach. We calculate total energies for a set of 
collinear spin configurations, and map these energies onto a classical Heisenberg model. Thus, total exchange integrals $J_i$ are estimated. In the LSDA+$U$ calculations, we use the Coulomb repulsion and exchange parameters $U_d=6.5\pm 1$~eV and $J_d=1$~eV, respectively.\cite{cu2v2o7,volborthite,janson2009} The double-counting-correction (DCC) scheme was set to the around-mean-field (AMF) option. The application of the fully-localized-limit (FLL) DCC had little effect on the exchange couplings.

\begin{figure}
\includegraphics{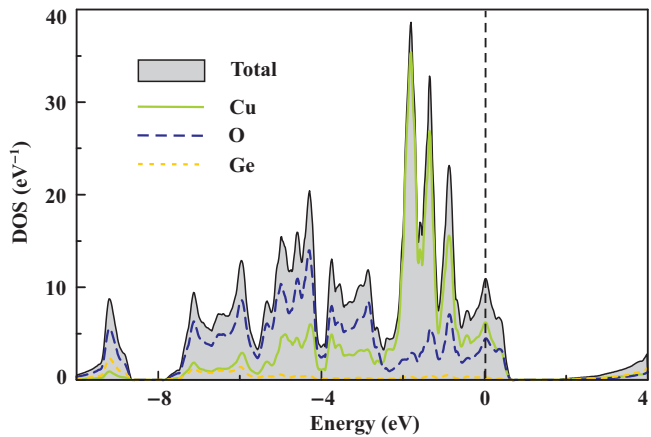}
\caption{\label{fig:dos}
(Color online) LDA density of states for Cu$_2$GeO$_4$. The Fermi level is at zero energy.
}
\end{figure}

The LDA energy spectrum of Cu$_2$GeO$_4$ is typical for Cu$^{+2}$ oxides. The mixed Cu $3d$ -- O $2p$ valence bands extend down to $-8$~eV (Fig.~\ref{fig:dos}), with the states near the Fermi level predominantly formed by the Cu $d_{x^2-y^2}$ orbital (here, $x$ and $y$ align with the short Cu--O bonds). Germanium orbitals contribute to the bands around $-10$~eV, and show negligible DOS at higher energies. While LDA yields a metallic energy spectrum due to the underestimation of electronic correlations in the Cu $3d$ shell, LSDA+$U$ restores the insulating scenario with the band gap of $E_g=2.0\pm 0.3$~eV for $U_d=6.5\pm 1$~eV. 

\begin{figure}
\includegraphics[scale=1]{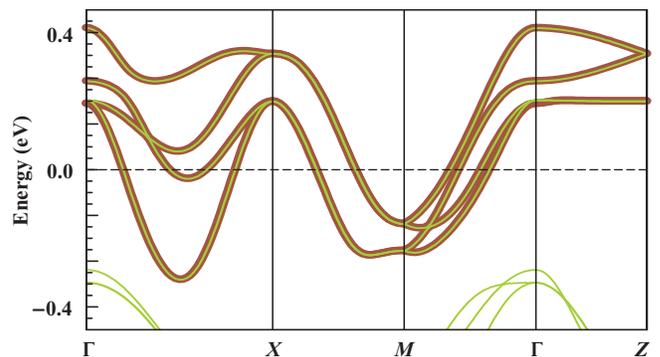}
\caption{\label{bands}
(Color online) LDA band structure of Cu$_2$GeO$_4$ (thin light lines) and the fit of the TB model (thick dark lines). The Fermi level is at zero energy. The $k$-path is defined as follows: $\Gamma(0,0,0)$, $X(0.5,0,0)$, $M(0.25,0.25,0)$, and $Z(0,0,0.5)$, where the coordinates are given in units of the reciprocal lattice parameters $4\pi/a$ and $4\pi/c$.
}
\end{figure}

\begin{table}
\caption{\label{exchanges}
Leading exchange couplings in Cu$_2$GeO$_4$: hopping parameters $t_i$ of the TB model, AFM contributions to the exchange couplings $J_i^{\AFM}=4t_i^2/U_{\eff}$, and the total exchange integrals $J_i$ from LSDA+$U$ calculations with $U_d=6.5$~eV.
}
\begin{ruledtabular}
\begin{tabular}{lcrrr}
           & Cu--Cu distance & $t_i$ & $J_i^{\AFM}$ & $J_i$  \\
           & (\r A)          & (meV) & (K)          & (K)    \\
  $J_1$    &  2.80           & 118   &  144         & $-60$  \\
  $J_2$    &  5.59           & 82    &   70         &   80   \\
  $J$      &  5.59        & 115   &  137         &  130  \\
  $J_{ab}$ &  6.25           & $-37$ &   14         &    7   \\
  $J_c$    &  3.07           & $-11$  &    1         &   $-2$ \\
\end{tabular}
\end{ruledtabular}
\end{table}
The Cu $d_{x^2-y^2}$ states are represented by four bands crossing the Fermi level and arising from four Cu atoms in the primitive cell of Cu$_2$GeO$_4$ (Fig.~\ref{bands}). These bands are separated from the rest of the valence bands by a pseudogap. To extract hopping 
parameters, we construct Wannier functions (WFs) based on the Cu $d_{x^2-y^2}$ 
character.\cite{wannier} This analysis evidences sizable nearest-neighbor ($t_1$) and next-nearest-neighbor ($t_2$) hoppings along the structural chains. However, the hopping $t$ along $a'$ is comparable to $t_1$ and $t_2$. Additionally, a weak diagonal hopping in the $ab$ plane is found (Table~\ref{exchanges}). The nearest-neighbor hoppings perpendicular to the $ab$ plane ($t_c$) are $-11$~meV, yielding $J_c^{\AFM}$ as low as $1$~K. The weak dispersion of the bands along $\Gamma-Z$ also shows the pronounced two-dimensionality of the system. Introducing the hoppings into an effective one-band Hubbard model, we evaluate AFM parts of the exchange integrals $J_i^{\AFM}$ (Table~\ref{exchanges}).

LSDA+$U$ calculations modify the LDA-based scenario. We find FM nearest-neighbor coupling within the structural chains, $J_1=-60\mp 10$~K for $U_d=6.5\pm 1$~eV. The next-nearest-neighbor intra-chain coupling $J_2=80\mp 20$~K and the interchain coupling $J=130\mp 30$~K are basically unchanged. Further couplings in the $ab$ plane are below 10~K. The interplane coupling becomes FM and remains weak. Thus, we establish the quasi-2D $J-J_1-J_2$ model with a weak interlayer coupling $J_c$ (Fig.~\ref{fig:structure}). 

The quasi-2D model of Cu$_2$GeO$_4$ results from the strong tetragonal distortion of the spinel structure. The plaquette description (Fig.~\ref{fig:structure}), with the magnetic $d_{x^2-y^2}$ orbital coplanar to the CuO$_4$ plaquette, clarifies the 2D nature of the system. The couplings $J_c$ connect the plaquettes lying in different planes, and therefore remain weak. By contrast, three sizable couplings in the $ab$ plane establish a frustrated spin lattice. Our model is dissimilar to the anisotropic pyrochlore lattice proposed by Yamada \textit{et al.}\cite{yamada2000} The pyrochlore spin lattice omits the relevant exchanges $J$ and $J_2$, and should be discarded. Cu$_2$GeO$_4$ is a frustrated magnet indeed, but the strong frustration is found in the $J_1-J_2$ chains rather than tetrahedral units.

The FM nearest-neighbor coupling $J_1$ should be referred to the Cu--O--Cu angle of $91.8^{\circ}$. The microscopic origin of ferromagnetism is the Hund's coupling on the oxygen site.\cite{mazurenko2007} The next-nearest-neighbor coupling $J_2$ is the AFM Cu--O--O--Cu superexchange. Similar values of $50-100$~K for $|J_1|$ and $J_2$ have been established for the archetype frustrated-spin-chain compounds, such as LiCu$_2$O$_2$ and 
LiCuVO$_4$.\cite{mazurenko2007,licuvo4,*comment} 

Another remark on the structural implementation of the spin model regards the origin of the long-range couplings $J$ and $J_2$. Since the Ge orbitals weakly contribute to the valence states, both couplings should be assigned to a Cu--O--O--Cu superexchange. Despite an identical Cu--Cu distance (Table~\ref{exchanges}), a larger $J$ value is caused by the co-planar arrangement of the plaquettes in the adjacent chains. By contrast, the next-nearest-neighbor plaquettes within the chain ($J_2$) lie in different planes due to the buckled chain geometry (Fig.~\ref{fig:structure}). It is worth to note that the $ab$ projections of the Cu$_2$GeO$_4$ and LiCuVO$_4$ structures are very similar. However, LiCuVO$_4$ is a quasi-1D system with $J\ll |J_1|,J_2$, while the spin system of Cu$_2$GeO$_4$ is quasi-2D.\cite{licuvo4,*comment,*enderle2010} 

\section{Microscopic model}
In the following, we explore the ground state and finite-temperature properties of our model. We first consider the purely 2D regime described by the Hamiltonian:
\begin{eqnarray}
\label{eq:model}
H&=& \sum_n \Big \{
\sum_i \big [
J_1\,{\bf s}_{i,n}\cdot{\bf s}_{i+1,n} + J_2\, {\bf s}_{i,n}\cdot{\bf
s}_{i+2,n}\big] \Big \}\nonumber\\
&+& \sum_i\sum_n J\, {\bf s}_{i,n}\cdot{\bf s}_{i,n+1} ,
\end{eqnarray}
where the index $n$ labels the structural chains (along $b'$), and $i$ denotes the lattice sites within a chain $n$. The effect of the interlayer coupling $J_c$ is discussed in Sec.~\ref{sec:tn}.

Our model can be viewed as frustrated $J_1-J_2$ chains (along $b'$) which are uniformly coupled by $J$ (along $a'$). Alternatively, one finds uniform spin chains along $a'$ with frustrated interchain couplings $J_1$ and $J_2$ along $b'$. While any of the parent 1D models is rather easy to handle, a rigorous treatment of their 2D combination is a challenging problem. Below, we apply the Lanczos diagonalization and coupled cluster method to achieve an accurate description of the ground state. By contrast, finite-temperature properties of the quantum model can only be accessed at high temperatures by a series expansion (HTSE), whereas conventional techniques, such as quantum Monte-Carlo or exact diagonalization, fail because of the sign problem or finite-size effects.\cite{note1}

\subsection{Magnetic susceptibility}
\label{sec:thermo}
\begin{figure}
\includegraphics{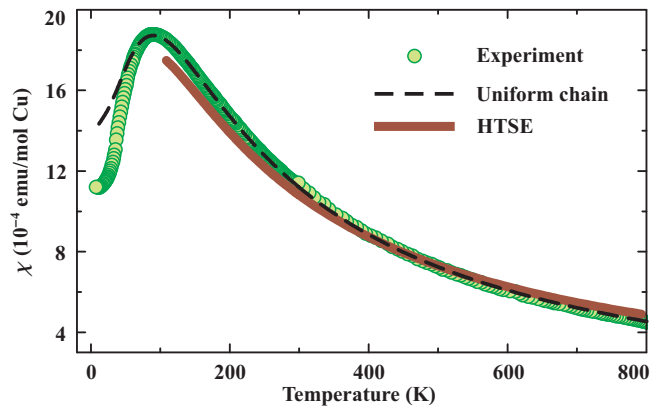}
\caption{\label{fig:chi}
(Color online) Fit of the experimental magnetic susceptibility data with the uniform chain model (dashed line) and the comparison to the HTSE of Eq.~\eqref{eq:htse} (solid line). Experimental data are from Ref.~\onlinecite{yamada2000}.
}
\end{figure}

Since the experimental information on Cu$_2$GeO$_4$ is restricted to the magnetic susceptibility and heat capacity measurements in Ref.~\onlinecite{yamada2000}, we discuss thermodynamic properties first. The experimental specific heat contains an unknown phonon contribution, hence the magnetic part can not be separated. Therefore, the magnetic susceptibility $\chi(T)$ (Fig.~\ref{fig:chi}) remains the only quantity suitable for the comparison between theory and experiment. The estimated Curie-Weiss temperature $\theta\!\simeq\! \frac12(J+J_1+J_2)\!=\!75$~K is in good agreement with the experimental value of $\theta=89$~K.\cite{yamada2000} 

For a further comparison, we derive the HTSE for our model:
\begin{equation}
  \chi=\dfrac{N_Ag^2\mu_B^2}{4k_B T}\left(1+\dfrac{J+J_1+J_2}{2T}+\dfrac{J^2+J_1^2+J_2^2}{4T^2}\right)^{-1},
\label{eq:htse}\end{equation}
where $N_A$ is Avogadro's number, $\mu_B$ is Bohr magneton, $g$ is the $g$-factor, and we used the expressions from Ref.~\onlinecite{johnston2000} up to the third order in temperature. The calculated exchange couplings (Table~\ref{exchanges}) are in reasonable agreement with the experimental data down to 150~K (Fig.~\ref{fig:chi}). The deviations at lower temperatures are likely related to the divergence of the HTSE at $T\leq J$. To improve the fit at higher temperatures, a slight adjustment of the exchange couplings is required. However, the third-order HTSE contains two $J$-dependent terms only, hence an unconstrained fit of three exchange parameters is impossible.

To access temperatures below 150~K, a simplification of the model is required. Since $J$ exceeds $|J_1|$ and $J_2$, the spin lattice is, to a first approximation, a set of uniform spin chains along $a'$. The respective 1D model\cite{johnston2000} fits the experimental magnetic susceptibility down to 70~K with $J=140$~K and $g=2.22$. A similar fit with $J=135$~K has been given in Ref.~\onlinecite{yamada2000}. However, Yamada~\textit{et al.}\cite{yamada2000} erroneously assign the spin chains to the structural chains (in their notation, $J$ corresponds to $J_1$). Our DFT calculations show that the uniform spin chains run perpendicular to the structural chains, whereas $J_1$ is FM. Such an intricate situation is not uncommon for low-dimensional magnets, see Refs.~\onlinecite{garrett1997} and~\onlinecite{kaul2003,*pb2v3o9} for similar examples.

Below 70~K, the 1D uniform-chain model overestimates the magnetic susceptibility of Cu$_2$GeO$_4$. This feature indicates an onset of 2D spin correlations. In contrast to the uniform spin-$\frac12$ chain having finite susceptibility at zero temperature, 2D and 3D systems usually develop a long-range order with vanishing susceptibility at low temperatures. The onset temperature of 2D spin correlations is a rough measure of interchain couplings. Indeed, the temperature of 70~K conforms to our estimates of $J_1=-60$~K and $J_2=80$~K, the couplings between the uniform spin chains.

\subsection{Coupled cluster method}
The coupled cluster method (CCM) and its application to frustrated spin systems have been previously reviewed in several articles, see, e.g., Refs.~\onlinecite{bursill,bishop98,krueger00,krueger01,farnell04,darradi_shastry,Schm:2006,darradi08,bishop08,zinke2009,bishop09,richter10,zinke10}.
Therefore, we will give only a brief illustration of the main relevant features of the method.
For more general information on the methodology of the CCM, see, e.g., Refs.~\onlinecite{bishop00,farnell04}, and references therein.

The CCM is a universal quantum many-body method. The starting point for a CCM calculation is the choice of a normalized reference or model state $|\Phi\rangle$, together with a complete set of (mutually commuting) multi-configurational creation operators $\{ C_L^+ \}$ and the corresponding set of their Hermitian adjoints $\{ C_L \}$. The CCM parametrizations of the ket- and bra- GSs are given by
\begin{eqnarray}
\label{eq_ccm}
|\Psi\rangle = e^S|\Phi\rangle, 
\qquad 
S = \sum_{I \neq 0}{\cal S}_IC_I^+ ; 
\nonumber\\
\langle\tilde{\Psi}| =  \langle\Phi|\tilde{S}e^{-S},
\qquad 
\tilde{S} = 1 + \sum_{I \neq 0}\tilde{\cal S}_IC_I .
\end{eqnarray}
Using $\langle\Phi|C_I^+=0=C_I|\Phi\rangle$ $\forall I\neq 0$, $C_0^+\equiv 1$, the commutation rules $[C_L^+,C_{K}^+] = 0=[C_L,C_{K}]$, the orthonormality condition $\langle\Phi|C_IC_J^+|\Phi\rangle=\delta_{IJ}$, and completeness $\displaystyle\sum_I C_I^+|\Phi\rangle\langle\Phi|C_I=1=|\Phi\rangle\langle\Phi|+\sum_{I\neq 0}C_I^+|\Phi\rangle\langle\Phi|C_I$, we get a set of non-linear and linear equations for the correlation coefficients ${\cal S}_I$ and $\tilde{\cal S}_I$, respectively. We choose a reference state corresponding to the classical state of the spin model, i.e., a non-collinear reference state with up-down \Neel-type correlations along the $a'$-direction (uniform $J$ chains) and with spiral
correlations along the $b'$-direction (frustrated $J_1-J_2$ chains). The spiral correlations are characterized by a pitch angle $\gamma$, i.e.\ $| \Phi\rangle = |\Phi(\gamma )\rangle $. In the quantum model, the pitch angle is typically different from the corresponding classical value $\gamma_{\rm cl}$. Hence, we do not choose the classical result for the pitch angle, and rather consider $\gamma$ as a free parameter in the CCM calculation. The value of $\gamma$ has to be determined by the minimization of the GS energy (in a certain CCM approximation, see below) given by $E(\gamma )= \langle \Phi (\gamma )| e^{-S}H e^S | \Phi(\gamma ) \rangle $, i.e., 
from the  $dE/d\gamma|_{\gamma=\gamma_{\rm qu}}=0$ condition.

In order to find an appropriate set of creation operators, it is convenient to perform a rotation of the local axes on each of the spins so that all spins in the reference state align with the negative $z$-direction. This rotation by an appropriate local angle $\delta_{i,n}=\delta_{i,n}(\gamma)$ of the spin on the lattice site $(i,n)$ is equivalent to the spin-operator transformation
\begin{equation}
\label{eq3} \left.\begin{array}{l} s_{i,n}^x = \cos\delta_{i,n} 
{\hat s}_{i,n}^x+\sin\delta_{i,n} {\hat s}_{i,n}^z; 
\quad s_{i,n}^y = {\hat s}_{i,n}^y  \\ s_{i,n}^z = 
-\sin\delta_{i,n} {\hat s}_{i,n}^x+\cos\delta_{i,n} {\hat s}_{i,n}^z 
\end{array} \right \} .
\end{equation}
The reference state and the corresponding creation operators $C_L^+$ are given by
\begin{equation}
 |{\hat \Phi}\ra = |\downarrow\downarrow\downarrow\downarrow\cdots\rangle \; ; \mbox{ } C_L^+ 
= {\hat s}_{i,n}^+ \, , \, {\hat s}_{i,n}^+{\hat s}_{j,m}^+ \, , \, 
{\hat s}_{i,n}^+{\hat s}_{j,m}^+{\hat s}_{k,l}^+ \, ,\, \ldots \; ,
\label{set1}
\end{equation}
where the indices $(i,n),(j,m),(k,l),\ldots$ denote arbitrary lattice sites. 
This specified form of the creation operators $C_L^+$ and the corresponding reference state $|{\hat \Phi}\ra$ immediately make clear that the general relations listed  below Eq.~(\ref{eq3}) are
fulfilled. In the rotated coordinate frame, the Heisenberg Hamiltonian acquires a dependence on the pitch angle $\gamma$ (see Ref.~\onlinecite{zinke2009} for more details).

The order parameter (sublattice magnetization) in the rotated coordinate frame is given by 
$m = -1/N \sum_{i,n}^N \langle\tilde\Psi|s_{i,n}^z|\Psi\rangle$. 
The only approximation of the CCM is the truncation of the expansion of the correlation operators $S$ and $\tilde S$. We use the well-established LSUB$n$ scheme, where all multispin correlations on the lattice with $n$ or fewer contiguous sites are taken into account.

In contrast to Ref.~\onlinecite{zinke2009}, which is focused on the $J\leq |J_1|,J_2$ regime, we consider the case of $J>|J_1|,J_2$. We also evaluate LSUB$n$ approximations of higher order, up to $n=8$. In the highest order of approximation, LSUB8, we have 21124 configurations, i.e., 21124 coupled non-linear equations have to be solved numerically. Moreover, the minimum of $E(\gamma)$ has to be found numerically to determine the quantum pitch angle $\gamma_{\rm qu}$. For the numerical calculations, we use the program package \texttt{CCCM} by D.J.J.~Farnell and J.~Schulenburg.\cite{cccm}

Since the LSUB$n$ becomes exact for $n \to \infty$, the numerical result can be improved 
by extrapolating the ``raw'' LSUB$n$ data to $n \to \infty$ using the
expression  $m_{n} = m_{\infty} + a/n + b/n^2$, cf. Refs.~\onlinecite{farnell04,krueger00,bishop00,bishop09}. 

\subsection{Ground state}
To find the classical ground state of the model given by Eq.~\eqref{eq:model}, we write the magnetic energy per lattice site for an arbitrary 2D propagation vector $\kv=(k_x,k_y)$:
\begin{equation}
  E=\frac12\left(J\cos k_x+J_1\cos k_y+J_2\cos(2k_y)\right),
\label{eq:energy2D}\end{equation}
where the unit cell of the spin lattice is used.\cite{note2} The energy minimum is found at $\kv=(\pi,\arccos(-\frac{J1}{4J_2}))$ that corresponds to the AFM order along $a'$ and spiral order along $b'$. The classical pitch angle is $\gamma=\arccos(-\frac{J1}{4J_2})=79.19^{\circ}$ and does not depend on $J$, hence the $a'$ and $b'$ directions of the spin lattice are fully decoupled. The ordering along $a'$ is controlled by $J$, whereas the ordering along $b'$ is controlled by the competing couplings $J_1$ and $J_2$. To check the validity of this result for the quantum case, we use the CCM method and Lanczos diagonalization.

\begin{figure}
\includegraphics{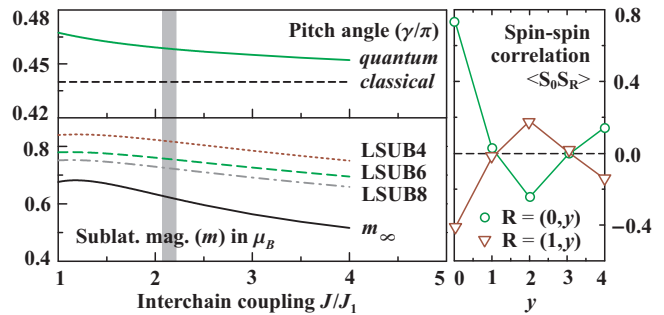}
\caption{\label{ccm}
(Color online) Left panel: the pitch angle ($\gamma$) and the sublattice magnetization
($m$) calculated by the CCM for $J_2=\frac{4}{3}|J_1|$. The shaded bar shows the coupling regime of Cu$_2$GeO$_4$. Note that the quantum pitch angles $\gamma_{\rm qu}$ for the LSUB$n$ approximations with $n=4,6$, and 8 almost coincide. Therefore, the shown LSUB8 curve represents effectively the limit $n\to\infty$. Right panel: spin-spin correlation $\langle {\bf S}_{0} {\bf S}_{\bf R}\rangle$, ${\bf R}=(x,y)$ along the $J_1-J_2$ chains ($b'$-axis) for a finite system of $N=32=4\times 8$ sites at $J_2=\frac{4}{3}|J_1|$ and $J=\frac{13}{6}|J_1|$.}
\end{figure}
In CCM, we reduce our exchange couplings (Table~\ref{exchanges}) to $J_1=-1$, $J_2=\frac43$, and vary $J$ (in Cu$_2$GeO$_4$, $J=\frac{13}{6}$). The CCM results for $\gamma$ and $m$ as a function of $J$ are shown in Fig.~\ref{ccm}. In contrast to the classical pitch angle $\gamma_{\rm cl}$, the pitch angle of the quantum system ($\gamma_{\rm qu}$) slightly depends on $J$. In Cu$_2$GeO$_4$, we find $\gamma_{\rm qu}= 83.9^{\circ}$, which is about $6\%$ larger than the classical angle, but about $5\%$ smaller than the quantum pitch angle for the isolated chain, i.e., at $J=0$. The coupling $J$ affects the dimensionality of the system and has a stronger effect on the sublattice magnetization (see Fig.~\ref{ccm}). The extrapolated value $m_\infty$ has a maximum at $J\simeq -1.17J_1$. The calculated exchange couplings in Cu$_2$GeO$_4$ lead to $m_\infty \sim 0.310$ (i.e., 0.62~$\mu_B$). 

The CCM results are confirmed by the Lanczos diagonalization data for the spin-spin correlation functions $\langle {\bf S}_{0} {\bf S}_{\bf R} \rangle$ shown in the right panel of Fig.~\ref{ccm}. We use a finite lattice comprising four 8-spin $J_1-J_2$ chains coupled by $J$. The correlations between nearest neighbors within the frustrated $J_1-J_2$ chains [$\mathbf R=(0,1)$] are close to zero, whereas the second-neighbor correlations [$\mathbf R=(0,2)$] are AFM. This conforms to the spiral ordering with the pitch angle close to $90^{\circ}$ (neighboring spins are nearly orthogonal). The correlations between the structural chains [at $\mathbf R=(1,y)$] follow the intra-chain correlations, yet showing the opposite sign. Thus, the ordering along $a'$ is AFM. 

\subsection{Long-range order}
\label{sec:tn}
The 2D model given by Eq.~\eqref{eq:model} is ordered at zero temperature only. To account for the actual long-range order in Cu$_2$GeO$_4$ below $T_N=33$~K, the interlayer coupling $J_c$ should be considered. The FM coupling $J_c$ is compatible with $J_1$, yet competing with $J$ and $J_2$. Assuming a similar ground state with the 2D propagation vector, we find that $J_c$ modifies the energy in Eq.~\eqref{eq:energy2D} by
\begin{equation}
  \Delta E_{\text{3D}}=\dfrac{J_c}{2}\cos\dfrac{k_x}{4}\cos\dfrac{k_y}{2}.
\end{equation}
Using DFT estimates of individual exchange couplings (Table~\ref{exchanges}), we arrive at the classical pitch angle modified by 0.15~\%: $79.07^{\circ}$ vs. $79.19^{\circ}$ for the purely 2D model. The classical energy per lattice site is reduced by 0.5~\% (about 0.6~K). This simplified analysis shows that the interlayer coupling $J_c$ is capable of stabilizing the 3D order in Cu$_2$GeO$_4$. However, the classical model does not reflect all the features of the real quantum model that, unfortunately, remains unfeasible for an accurate numerical study. In particular, the ordering temperature $T_N$ can not be determined with sufficient accuracy.

\section{Discussion and summary}
Although not obvious at first glance, the microscopic magnetic model of Cu$_2$GeO$_4$ can be deduced from simple qualitative arguments. While the naive geometrical analysis of the crystal structure suggests a 3D pyrochlore-lattice magnetism, a closer look at the crystal structure identifies 2D features. In most of the Cu$^{+2}$ oxides, electronic structure and magnetism are controlled by the arrangement of CuO$_4$ plaquettes, which are the basic structural entities. Chains of edge-sharding plaquettes give rise to frustrated $J_1-J_2$ spin chains,\cite{gippius2004,*masuda2005,capogna2005,*drechsler2006,licuvo4,*comment,*enderle2010,cucl2,*banks2009} yet the coplanar arrangement of the plaquettes in the neighboring structural chains induces a strong AFM coupling along $a'$. Overall, we find magnetic layers in the $ab$ plane, along with a weak FM interlayer coupling $J_c$. By combining the frustration along $b'$ with the strong unfrustrated exchange along $a'$, Cu$_2$GeO$_4$ expands the family of cuprates featuring frustrated spin chains. 

The dearth of the experimental data and the complexity of the 2D frustrated $J-J_1-J_2$ lattice restrict the opportunities for an experimental verification of our microscopic model. Nevertheless, the tangible success of DFT in unraveling complex spin lattices for a range of transition-metal compounds\cite{dioptase,kaul2003,*pb2v3o9,valenti2003,cucl} is a solid justification of our results. The 2D nature of the system and the frustrated couplings along $b'$ are confirmed by qualitative arguments and by a reference to similar Cu$^{+2}$ compounds (see Sec.~\ref{sec:band}). Further on, numerical estimates of individual exchanges conform to the experimental magnetic susceptibility (Sec.~\ref{sec:thermo}). An ultimate test of the proposed model requires a study of the ground state by neutron or resonant x-ray scattering. Presently, we note that our model does predict the long-range magnetic order, in contrast to the strongly anisotropic pyrochlore lattice that might have a quantum-disordered valence-bond-solid\cite{starykh2005} or a gapless spin-liquid\cite{sindzingre2002} ground state.

The experimental data for Cu$_2$GeO$_4$ and accurate theoretical results for the ground state disclose the basic features of our model. At high temperatures, thermodynamic properties are guided by the uniform spin chains along $a'$. The apparent spin-chain behavior is likely related to the partial cancellation of FM $J_1$ and AFM $J_2$ in the second-order term for the susceptibility (see Eq.~\eqref{eq:htse}). A further evidence is the perfect fit of the experimental magnetic susceptibility down to 70~K ($T/J\simeq 0.5$). The frustrated couplings along $b'$ come into play at lower temperatures, and essentially determine the ground state. The leading exchange $J$ drives AFM ordering along $a'$. The ordering along $b'$ has to satisfy the frustrated couplings $J_1$ and $J_2$ and is, therefore, a spiral, similar to a single $J_1-J_2$ frustrated spin chain. The interlayer coupling $J_c$ should stabilize the long-range order up to $T_N$ without changing the basic features of the ground state: the collinear AFM order along $a'$ and the spiral order along $b'$.

The highly accurate CCM approach provides reliable information on the ground state of the 2D system. We find the pitch angle of $\gamma\simeq 84^{\circ}$ and the sublattice magnetization close to 0.62~$\mu_B$. Both $\gamma$ and $m$ are renormalized with respect to the classical values, and suggest strong quantum fluctuations in the system. Enhanced quantum fluctuations should be ascribed to the reduced dimensionality and frustration. The magnetic ordering temperature $T_N/J\simeq 0.25$ is also suggestive of strong quantum fluctuations. For example, a quasi-2D system of square lattices with a weak interlayer coupling $J_{\perp}/J=0.01$ (compare to $|J_c|/J\simeq 0.01$) orders at a higher temperature of $T_N/J\simeq 0.33$ (Ref.~\onlinecite{yasuda2005}). The quantum effects in the system could be further probed by an experimental study of the ground state.

An interesting feature of the spiral magnetic order is the possible emergence of electric polarization strongly coupled to the magnetism.\cite{mostovoy} The direction of the electric polarization depends on the twisting direction of the spiral. In Cu$_2$GeO$_4$, the AFM coupling $J$ leads to opposite twisting directions in the neighboring spirals, hence the polarization is canceled. However, the proposed antiferroelectricity of Cu$_2$GeO$_4$ does not preclude the strong magnetoelectric coupling, and should stimulate further experimental investigation of the compound. We also note that the family of 2D frustrated materials representing the $J-J_1-J_2$ model can be further expanded by CuNCN lying in the limit of $J\gg |J_1|,J_2$.\cite{tsirlin2010}
 
In summary, we have shown that the electronic structure of Cu$_2$GeO$_4$ contradicts the previous, empirical-based spin model of the anisotropic pyrochlore lattice. The comprehensive computational study discloses the quasi-2D nature of this compound and suggests an original 2D spin model comprising frustrated and uniform spin chains along the two dimensions. Theoretical results for this model show the robust nature of the spiral ground state that is subject to strong quantum effects, evidenced by the reduced sublattice magnetization of 0.62~$\mu_B$ and the renormalized pitch angle of about $84^{\circ}$. 

\acknowledgments
We are grateful to Oleg Janson and Deepa Kasinathan for fruitful discussions and careful reading of the manuscript. A. T. acknowledges financial support from Alexander von Humboldt Foundation. J. R. appreciates the funding by DFG (project RI 615/16-1).


\begin{thebibliography}{64}%
\makeatletter
\providecommand \@ifxundefined [1]{%
 \@ifx{#1\undefined}
}%
\providecommand \@ifnum [1]{%
 \ifnum #1\expandafter \@firstoftwo
 \else \expandafter \@secondoftwo
 \fi
}%
\providecommand \@ifx [1]{%
 \ifx #1\expandafter \@firstoftwo
 \else \expandafter \@secondoftwo
 \fi
}%
\providecommand \natexlab [1]{#1}%
\providecommand \enquote  [1]{``#1''}%
\providecommand \bibnamefont  [1]{#1}%
\providecommand \bibfnamefont [1]{#1}%
\providecommand \citenamefont [1]{#1}%
\providecommand \href@noop [0]{\@secondoftwo}%
\providecommand \href [0]{\begingroup \@sanitize@url \@href}%
\providecommand \@href[1]{\@@startlink{#1}\@@href}%
\providecommand \@@href[1]{\endgroup#1\@@endlink}%
\providecommand \@sanitize@url [0]{\catcode `\\12\catcode `\$12\catcode
  `\&12\catcode `\#12\catcode `\^12\catcode `\_12\catcode `\%12\relax}%
\providecommand \@@startlink[1]{}%
\providecommand \@@endlink[0]{}%
\providecommand \url  [0]{\begingroup\@sanitize@url \@url }%
\providecommand \@url [1]{\endgroup\@href {#1}{\urlprefix }}%
\providecommand \urlprefix  [0]{URL }%
\providecommand \Eprint [0]{\href }%
\@ifxundefined \urlstyle {%
  \providecommand \doi  [0]{\begingroup \@sanitize@url \@doi}%
  \providecommand \@doi [1]{\endgroup \@@startlink {\doibase
  #1}doi:\discretionary {}{}{}#1\@@endlink }%
}{%
  \providecommand \doi  [0]{doi:\discretionary{}{}{}\begingroup
  \urlstyle{rm}\Url }%
}%
\providecommand \doibase [0]{http://dx.doi.org/}%
\providecommand \Doi [0]{\begingroup \@sanitize@url \@Doi }%
\providecommand \@Doi  [1]{\endgroup\@@startlink{\doibase#1}\@@Doi}%
\providecommand \@@Doi [1]{#1\@@endlink}%
\providecommand \selectlanguage [0]{\@gobble}%
\providecommand \bibinfo  [0]{\@secondoftwo}%
\providecommand \bibfield  [0]{\@secondoftwo}%
\providecommand \translation [1]{[#1]}%
\providecommand \BibitemOpen [0]{}%
\providecommand \bibitemStop [0]{}%
\providecommand \bibitemNoStop [0]{.\EOS\space}%
\providecommand \EOS [0]{\spacefactor3000\relax}%
\providecommand \BibitemShut  [1]{\csname bibitem#1\endcsname}%
\bibitem [{\citenamefont {Diep}(2004)}]{frustrated}%
  \BibitemOpen
  \bibinfo {editor} {\bibfnamefont {H.~T.}\ \bibnamefont {Diep}},\ ed.,\
  \href@noop {} {\emph {\bibinfo {title} {Frustrated spin systems}}}\ (\bibinfo
   {publisher} {World Scientific},\ \bibinfo {year} {2004})\BibitemShut
  {NoStop}%
\bibitem [{\citenamefont {Schollw\"ock}\ \emph {et~al.}(2004)\citenamefont
  {Schollw\"ock}, \citenamefont {Richter}, \citenamefont {Farnell},\ and\
  \citenamefont {Bishop}}]{quantum}%
  \BibitemOpen
  \bibinfo {editor} {\bibfnamefont {U.}~\bibnamefont {Schollw\"ock}}, \bibinfo
  {editor} {\bibfnamefont {J.}~\bibnamefont {Richter}}, \bibinfo {editor}
  {\bibfnamefont {D.~J.~J.}\ \bibnamefont {Farnell}}, \ and\ \bibinfo {editor}
  {\bibfnamefont {R.~F.}\ \bibnamefont {Bishop}},\ eds.,\ \href@noop {} {\emph
  {\bibinfo {title} {Quantum magnetism}}}\ (\bibinfo  {publisher} {Springer},\
  \bibinfo {year} {2004})\BibitemShut {NoStop}%
\bibitem [{\citenamefont {Sologubenko}\ \emph {et~al.}(2007)\citenamefont
  {Sologubenko}, \citenamefont {Lorenz}, \citenamefont {Ott},\ and\
  \citenamefont {Freimuth}}]{ballistic}%
  \BibitemOpen
  \bibfield  {author} {\bibinfo {author} {\bibfnamefont {A.~V.}\ \bibnamefont
  {Sologubenko}}, \bibinfo {author} {\bibfnamefont {T.}~\bibnamefont {Lorenz}},
  \bibinfo {author} {\bibfnamefont {H.~R.}\ \bibnamefont {Ott}}, \ and\
  \bibinfo {author} {\bibfnamefont {A.}~\bibnamefont {Freimuth}},\ }\href@noop
  {} {\bibfield  {journal} {\bibinfo  {journal} {J. Low-Temp. Phys.},\ }\textbf
  {\bibinfo {volume} {147}},\ \bibinfo {pages} {387} (\bibinfo {year}
  {2007}), cond-mat/0611052}\BibitemShut {NoStop}%
\bibitem [{\citenamefont {Zhitomirsky}\ and\ \citenamefont
  {Honecker}(2004)}]{zhitomirsky2004}%
  \BibitemOpen
  \bibfield  {author} {\bibinfo {author} {\bibfnamefont {M.~E.}\ \bibnamefont
  {Zhitomirsky}}\ and\ \bibinfo {author} {\bibfnamefont {A.}~\bibnamefont
  {Honecker}},\ }\href@noop {} {\bibfield  {journal} {\bibinfo  {journal} {J.
  Stat. Mech.},\ \bibinfo {pages} {P07012}} (\bibinfo {year}
  {2004}), cond-mat/0404683}\BibitemShut {NoStop}%
\bibitem [{\citenamefont {Schnack}\ \emph {et~al.}(2007)\citenamefont
  {Schnack}, \citenamefont {Schmidt},\ and\ \citenamefont
  {Richter}}]{schnack2007}%
  \BibitemOpen
  \bibfield  {author} {\bibinfo {author} {\bibfnamefont {J.}~\bibnamefont
  {Schnack}}, \bibinfo {author} {\bibfnamefont {R.}~\bibnamefont {Schmidt}}, \
  and\ \bibinfo {author} {\bibfnamefont {J.}~\bibnamefont {Richter}},\
  }\href@noop {} {\bibfield  {journal} {\bibinfo  {journal} {Phys. Rev. B},\
  }\textbf {\bibinfo {volume} {76}},\ \bibinfo {pages} {054413} (\bibinfo
  {year} {2007}), cond-mat/0703480}\BibitemShut {NoStop}%
\bibitem [{\citenamefont {Cheong}\ and\ \citenamefont
  {Mostovoy}(2007)}]{mostovoy}%
  \BibitemOpen
  \bibfield  {author} {\bibinfo {author} {\bibfnamefont {S.-W.}\ \bibnamefont
  {Cheong}}\ and\ \bibinfo {author} {\bibfnamefont {M.}~\bibnamefont
  {Mostovoy}},\ }\href@noop {} {\bibfield  {journal} {\bibinfo  {journal}
  {Nature Materials},\ }\textbf {\bibinfo {volume} {6}},\ \bibinfo {pages} {13}
  (\bibinfo {year} {2007})}\BibitemShut {NoStop}%
\bibitem [{\citenamefont {Furukawa}\ \emph {et~al.}(2010)\citenamefont
  {Furukawa}, \citenamefont {Sato},\ and\ \citenamefont
  {Onoda}}]{furukawa2010}%
  \BibitemOpen
  \bibfield  {author} {\bibinfo {author} {\bibfnamefont {S.}~\bibnamefont
  {Furukawa}}, \bibinfo {author} {\bibfnamefont {M.}~\bibnamefont {Sato}}, \
  and\ \bibinfo {author} {\bibfnamefont {S.}~\bibnamefont {Onoda}},\
  }\href@noop {} {\bibfield  {journal} {\bibinfo  {journal} {Phys. Rev.
  Lett.},\ }\textbf {\bibinfo {volume} {105}},\ \bibinfo {pages} {257205}
  (\bibinfo {year} {2010}), arXiv:1003.3940}\BibitemShut {NoStop}%
\bibitem [{\citenamefont {Mazurenko}\ \emph {et~al.}(2007)\citenamefont
  {Mazurenko}, \citenamefont {Skornyakov}, \citenamefont {Kozhevnikov},
  \citenamefont {Mila},\ and\ \citenamefont {Anisimov}}]{mazurenko2007}%
  \BibitemOpen
  \bibfield  {author} {\bibinfo {author} {\bibfnamefont {V.~V.}\ \bibnamefont
  {Mazurenko}}, \bibinfo {author} {\bibfnamefont {S.~L.}\ \bibnamefont
  {Skornyakov}}, \bibinfo {author} {\bibfnamefont {A.~V.}\ \bibnamefont
  {Kozhevnikov}}, \bibinfo {author} {\bibfnamefont {F.}~\bibnamefont {Mila}}, \
  and\ \bibinfo {author} {\bibfnamefont {V.~I.}\ \bibnamefont {Anisimov}},\
  }\href@noop {} {\bibfield  {journal} {\bibinfo  {journal} {Phys. Rev. B},\
  }\textbf {\bibinfo {volume} {75}},\ \bibinfo {pages} {224408} (\bibinfo
  {year} {2007})},\ \bibinfo {note} {and references therein}, cond-mat/0702276\BibitemShut
  {NoStop}%
\bibitem [{\citenamefont {Gippius}\ \emph {et~al.}(2004)\citenamefont
  {Gippius}, \citenamefont {Morozova}, \citenamefont {Moskvin}, \citenamefont
  {Zalessky}, \citenamefont {Bush}, \citenamefont {Baenitz}, \citenamefont
  {Rosner},\ and\ \citenamefont {Drechsler}}]{gippius2004}%
  \BibitemOpen
  \bibfield  {author} {\bibinfo {author} {\bibfnamefont {A.~A.}\ \bibnamefont
  {Gippius}}, \bibinfo {author} {\bibfnamefont {E.~N.}\ \bibnamefont
  {Morozova}}, \bibinfo {author} {\bibfnamefont {A.~S.}\ \bibnamefont
  {Moskvin}}, \bibinfo {author} {\bibfnamefont {A.~V.}\ \bibnamefont
  {Zalessky}}, \bibinfo {author} {\bibfnamefont {A.~A.}\ \bibnamefont {Bush}},
  \bibinfo {author} {\bibfnamefont {M.}~\bibnamefont {Baenitz}}, \bibinfo
  {author} {\bibfnamefont {H.}~\bibnamefont {Rosner}}, \ and\ \bibinfo {author}
  {\bibfnamefont {S.-L.}\ \bibnamefont {Drechsler}},\ }\href@noop {} {\bibfield
   {journal} {\bibinfo  {journal} {Phys. Rev. B},\ }\textbf {\bibinfo {volume}
  {70}},\ \bibinfo {pages} {020406(R)} (\bibinfo {year} {2004}), cond-mat/0312706}\BibitemShut
  {NoStop}%
\bibitem [{\citenamefont {Masuda}\ \emph {et~al.}(2005)\citenamefont {Masuda},
  \citenamefont {Zheludev}, \citenamefont {Roessli}, \citenamefont {Bush},
  \citenamefont {Markina},\ and\ \citenamefont {Vasiliev}}]{masuda2005}%
  \BibitemOpen
  \bibfield  {author} {\bibinfo {author} {\bibfnamefont {T.}~\bibnamefont
  {Masuda}}, \bibinfo {author} {\bibfnamefont {A.}~\bibnamefont {Zheludev}},
  \bibinfo {author} {\bibfnamefont {B.}~\bibnamefont {Roessli}}, \bibinfo
  {author} {\bibfnamefont {A.}~\bibnamefont {Bush}}, \bibinfo {author}
  {\bibfnamefont {M.}~\bibnamefont {Markina}}, \ and\ \bibinfo {author}
  {\bibfnamefont {A.}~\bibnamefont {Vasiliev}},\ }\href@noop {} {\bibfield
  {journal} {\bibinfo  {journal} {Phys. Rev. B},\ }\textbf {\bibinfo {volume}
  {72}},\ \bibinfo {pages} {014405} (\bibinfo {year} {2005}), cond-mat/0412625}\BibitemShut
  {NoStop}%
\bibitem [{\citenamefont {Enderle}\ \emph {et~al.}(2005)\citenamefont
  {Enderle}, \citenamefont {Mukherjee}, \citenamefont {F{\r a}k}, \citenamefont
  {Kremer}, \citenamefont {Broto}, \citenamefont {Rosner}, \citenamefont
  {Drechsler}, \citenamefont {Richter}, \citenamefont {Malek}, \citenamefont
  {Prokofiev}, \citenamefont {Assmus}, \citenamefont {Pujol}, \citenamefont
  {Raggazzoni}, \citenamefont {Rakoto}, \citenamefont {Rheinst\"adter},\ and\
  \citenamefont {R{\o n}now}}]{licuvo4}%
  \BibitemOpen
  \bibfield  {author} {\bibinfo {author} {\bibfnamefont {M.}~\bibnamefont
  {Enderle}}, \bibinfo {author} {\bibfnamefont {C.}~\bibnamefont {Mukherjee}},
  \bibinfo {author} {\bibfnamefont {B.}~\bibnamefont {F{\r a}k}}, \bibinfo
  {author} {\bibfnamefont {R.~K.}\ \bibnamefont {Kremer}}, \bibinfo {author}
  {\bibfnamefont {J.-M.}\ \bibnamefont {Broto}}, \bibinfo {author}
  {\bibfnamefont {H.}~\bibnamefont {Rosner}}, \bibinfo {author} {\bibfnamefont
  {S.-L.}\ \bibnamefont {Drechsler}}, \bibinfo {author} {\bibfnamefont
  {J.}~\bibnamefont {Richter}}, \bibinfo {author} {\bibfnamefont
  {J.}~\bibnamefont {Malek}}, \bibinfo {author} {\bibfnamefont
  {A.}~\bibnamefont {Prokofiev}}, \bibinfo {author} {\bibfnamefont
  {W.}~\bibnamefont {Assmus}}, \bibinfo {author} {\bibfnamefont
  {S.}~\bibnamefont {Pujol}}, \bibinfo {author} {\bibfnamefont {J.-L.}\
  \bibnamefont {Raggazzoni}}, \bibinfo {author} {\bibfnamefont
  {H.}~\bibnamefont {Rakoto}}, \bibinfo {author} {\bibfnamefont
  {M.}~\bibnamefont {Rheinst\"adter}}, \ and\ \bibinfo {author} {\bibfnamefont
  {H.~M.}\ \bibnamefont {R{\o n}now}},\ }\href@noop {} {\bibfield  {journal}
  {\bibinfo  {journal} {Europhys. Lett.},\ }\textbf {\bibinfo {volume} {70}},\
  \bibinfo {pages} {237} (\bibinfo {year} {2005})}\BibitemShut {NoStop}%
\bibitem [{\citenamefont {Drechsler}\ \emph {et~al.}(2010)\citenamefont
  {Drechsler}, \citenamefont {Nishimoto}, \citenamefont {Kuzian}, \citenamefont
  {M\'alek}, \citenamefont {Richter}, \citenamefont {van~den Brink},
  \citenamefont {Schmitt},\ and\ \citenamefont {Rosner}}]{comment}%
  \BibitemOpen
  \bibfield  {author} {\bibinfo {author} {\bibfnamefont {S.-L.}\ \bibnamefont
  {Drechsler}}, \bibinfo {author} {\bibfnamefont {S.}~\bibnamefont
  {Nishimoto}}, \bibinfo {author} {\bibfnamefont {R.}~\bibnamefont {Kuzian}},
  \bibinfo {author} {\bibfnamefont {J.}~\bibnamefont {M\'alek}}, \bibinfo
  {author} {\bibfnamefont {J.}~\bibnamefont {Richter}}, \bibinfo {author}
  {\bibfnamefont {J.}~\bibnamefont {van~den Brink}}, \bibinfo {author}
  {\bibfnamefont {M.}~\bibnamefont {Schmitt}}, \ and\ \bibinfo {author}
  {\bibfnamefont {H.}~\bibnamefont {Rosner}},\ }\href@noop {} {} (\bibinfo
  {year} {2010}),\ \Eprint {http://arxiv.org/abs/1006.5070} {arXiv:1006.5070}
  \BibitemShut {NoStop}%
\bibitem [{\citenamefont {Enderle}\ \emph {et~al.}(2010)\citenamefont
  {Enderle}, \citenamefont {F{\r a}k}, \citenamefont {Mikeska}, \citenamefont
  {Kremer}, \citenamefont {Prokofiev},\ and\ \citenamefont
  {Assmus}}]{enderle2010}%
  \BibitemOpen
  \bibfield  {author} {\bibinfo {author} {\bibfnamefont {M.}~\bibnamefont
  {Enderle}}, \bibinfo {author} {\bibfnamefont {B.}~\bibnamefont {F{\r a}k}},
  \bibinfo {author} {\bibfnamefont {H.-J.}\ \bibnamefont {Mikeska}}, \bibinfo
  {author} {\bibfnamefont {R.~K.}\ \bibnamefont {Kremer}}, \bibinfo {author}
  {\bibfnamefont {A.}~\bibnamefont {Prokofiev}}, \ and\ \bibinfo {author}
  {\bibfnamefont {W.}~\bibnamefont {Assmus}},\ }\href@noop {} {\bibfield
  {journal} {\bibinfo  {journal} {Phys. Rev. Lett.},\ }\textbf {\bibinfo
  {volume} {104}},\ \bibinfo {pages} {237207} (\bibinfo {year}
  {2010})}\BibitemShut {NoStop}%
\bibitem [{\citenamefont {Capogna}\ \emph {et~al.}(2005)\citenamefont
  {Capogna}, \citenamefont {Mayr}, \citenamefont {Horsch}, \citenamefont
  {Raichle}, \citenamefont {Kremer}, \citenamefont {Sofin}, \citenamefont
  {Maljuk}, \citenamefont {Jansen},\ and\ \citenamefont
  {Keimer}}]{capogna2005}%
  \BibitemOpen
  \bibfield  {author} {\bibinfo {author} {\bibfnamefont {L.}~\bibnamefont
  {Capogna}}, \bibinfo {author} {\bibfnamefont {M.}~\bibnamefont {Mayr}},
  \bibinfo {author} {\bibfnamefont {P.}~\bibnamefont {Horsch}}, \bibinfo
  {author} {\bibfnamefont {M.}~\bibnamefont {Raichle}}, \bibinfo {author}
  {\bibfnamefont {R.~K.}\ \bibnamefont {Kremer}}, \bibinfo {author}
  {\bibfnamefont {M.}~\bibnamefont {Sofin}}, \bibinfo {author} {\bibfnamefont
  {A.}~\bibnamefont {Maljuk}}, \bibinfo {author} {\bibfnamefont
  {M.}~\bibnamefont {Jansen}}, \ and\ \bibinfo {author} {\bibfnamefont
  {B.}~\bibnamefont {Keimer}},\ }\href@noop {} {\bibfield  {journal} {\bibinfo
  {journal} {Phys. Rev. B},\ }\textbf {\bibinfo {volume} {71}},\ \bibinfo
  {pages} {140402(R)} (\bibinfo {year} {2005}), cond-mat/0411753}\BibitemShut {NoStop}%
\bibitem [{\citenamefont {Drechsler}\ \emph {et~al.}(2006)\citenamefont
  {Drechsler}, \citenamefont {Richter}, \citenamefont {Gippius}, \citenamefont
  {Vasiliev}, \citenamefont {Bush}, \citenamefont {Moskvin}, \citenamefont
  {M\'alek}, \citenamefont {Prots}, \citenamefont {Schnelle},\ and\
  \citenamefont {Rosner}}]{drechsler2006}%
  \BibitemOpen
  \bibfield  {author} {\bibinfo {author} {\bibfnamefont {S.-L.}\ \bibnamefont
  {Drechsler}}, \bibinfo {author} {\bibfnamefont {J.}~\bibnamefont {Richter}},
  \bibinfo {author} {\bibfnamefont {A.~A.}\ \bibnamefont {Gippius}}, \bibinfo
  {author} {\bibfnamefont {A.}~\bibnamefont {Vasiliev}}, \bibinfo {author}
  {\bibfnamefont {A.~A.}\ \bibnamefont {Bush}}, \bibinfo {author}
  {\bibfnamefont {A.~S.}\ \bibnamefont {Moskvin}}, \bibinfo {author}
  {\bibfnamefont {J.}~\bibnamefont {M\'alek}}, \bibinfo {author} {\bibfnamefont
  {Y.}~\bibnamefont {Prots}}, \bibinfo {author} {\bibfnamefont
  {W.}~\bibnamefont {Schnelle}}, \ and\ \bibinfo {author} {\bibfnamefont
  {H.}~\bibnamefont {Rosner}},\ }\href@noop {} {\bibfield  {journal} {\bibinfo
  {journal} {Europhys. Lett.},\ }\textbf {\bibinfo {volume} {73}},\ \bibinfo
  {pages} {83} (\bibinfo {year} {2006})}\BibitemShut {NoStop}%
\bibitem [{\citenamefont {Schmitt}\ \emph {et~al.}(2009)\citenamefont
  {Schmitt}, \citenamefont {Janson}, \citenamefont {Schmidt}, \citenamefont
  {Hoffmann}, \citenamefont {Schnelle}, \citenamefont {Drechsler},\ and\
  \citenamefont {Rosner}}]{cucl2}%
  \BibitemOpen
  \bibfield  {author} {\bibinfo {author} {\bibfnamefont {M.}~\bibnamefont
  {Schmitt}}, \bibinfo {author} {\bibfnamefont {O.}~\bibnamefont {Janson}},
  \bibinfo {author} {\bibfnamefont {M.}~\bibnamefont {Schmidt}}, \bibinfo
  {author} {\bibfnamefont {S.}~\bibnamefont {Hoffmann}}, \bibinfo {author}
  {\bibfnamefont {W.}~\bibnamefont {Schnelle}}, \bibinfo {author}
  {\bibfnamefont {S.-L.}\ \bibnamefont {Drechsler}}, \ and\ \bibinfo {author}
  {\bibfnamefont {H.}~\bibnamefont {Rosner}},\ }\href@noop {} {\bibfield
  {journal} {\bibinfo  {journal} {Phys. Rev. B},\ }\textbf {\bibinfo {volume}
  {79}},\ \bibinfo {pages} {245119} (\bibinfo {year} {2009}), arXiv:0905.4038}\BibitemShut
  {NoStop}%
\bibitem [{\citenamefont {Banks}\ \emph {et~al.}(2009)\citenamefont {Banks},
  \citenamefont {Kremer}, \citenamefont {Hoch}, \citenamefont {Simon},
  \citenamefont {Ouladdiaf}, \citenamefont {Broto}, \citenamefont {Rakoto},
  \citenamefont {Lee},\ and\ \citenamefont {Whangbo}}]{banks2009}%
  \BibitemOpen
  \bibfield  {author} {\bibinfo {author} {\bibfnamefont {M.~G.}\ \bibnamefont
  {Banks}}, \bibinfo {author} {\bibfnamefont {R.~K.}\ \bibnamefont {Kremer}},
  \bibinfo {author} {\bibfnamefont {C.}~\bibnamefont {Hoch}}, \bibinfo {author}
  {\bibfnamefont {A.}~\bibnamefont {Simon}}, \bibinfo {author} {\bibfnamefont
  {B.}~\bibnamefont {Ouladdiaf}}, \bibinfo {author} {\bibfnamefont {J.-M.}\
  \bibnamefont {Broto}}, \bibinfo {author} {\bibfnamefont {H.}~\bibnamefont
  {Rakoto}}, \bibinfo {author} {\bibfnamefont {C.}~\bibnamefont {Lee}}, \ and\
  \bibinfo {author} {\bibfnamefont {M.-H.}\ \bibnamefont {Whangbo}},\
  }\href@noop {} {\bibfield  {journal} {\bibinfo  {journal} {Phys. Rev. B},\
  }\textbf {\bibinfo {volume} {80}},\ \bibinfo {pages} {024404} (\bibinfo
  {year} {2009}), arXiv:0904.2929}\BibitemShut {NoStop}%
\bibitem [{\citenamefont {Tsirlin}\ and\ \citenamefont
  {Rosner}(2010){\natexlab{a}}}]{tsirlin2010}%
  \BibitemOpen
  \bibfield  {author} {\bibinfo {author} {\bibfnamefont {A.~A.}\ \bibnamefont
  {Tsirlin}}\ and\ \bibinfo {author} {\bibfnamefont {H.}~\bibnamefont
  {Rosner}},\ }\href@noop {} {\bibfield  {journal} {\bibinfo  {journal} {Phys.
  Rev. B},\ }\textbf {\bibinfo {volume} {81}},\ \bibinfo {pages} {024424}
  (\bibinfo {year} {2010}{\natexlab{a}}), arXiv:0910.2056}\BibitemShut {NoStop}%
\bibitem [{\citenamefont {Park}\ \emph {et~al.}(2007)\citenamefont {Park},
  \citenamefont {Choi}, \citenamefont {Zhang},\ and\ \citenamefont
  {Cheong}}]{park2007}%
  \BibitemOpen
  \bibfield  {author} {\bibinfo {author} {\bibfnamefont {S.}~\bibnamefont
  {Park}}, \bibinfo {author} {\bibfnamefont {Y.~J.}\ \bibnamefont {Choi}},
  \bibinfo {author} {\bibfnamefont {C.~L.}\ \bibnamefont {Zhang}}, \ and\
  \bibinfo {author} {\bibfnamefont {S.-W.}\ \bibnamefont {Cheong}},\
  }\href@noop {} {\bibfield  {journal} {\bibinfo  {journal} {Phys. Rev.
  Lett.},\ }\textbf {\bibinfo {volume} {98}},\ \bibinfo {pages} {057601}
  (\bibinfo {year} {2007})}\BibitemShut {NoStop}%
\bibitem [{\citenamefont {Naito}\ \emph {et~al.}(2007)\citenamefont {Naito},
  \citenamefont {Sato}, \citenamefont {Yasui}, \citenamefont {Kobayashi},
  \citenamefont {Kobayashi},\ and\ \citenamefont {Sato}}]{naito2007}%
  \BibitemOpen
  \bibfield  {author} {\bibinfo {author} {\bibfnamefont {Y.}~\bibnamefont
  {Naito}}, \bibinfo {author} {\bibfnamefont {K.}~\bibnamefont {Sato}},
  \bibinfo {author} {\bibfnamefont {Y.}~\bibnamefont {Yasui}}, \bibinfo
  {author} {\bibfnamefont {Y.}~\bibnamefont {Kobayashi}}, \bibinfo {author}
  {\bibfnamefont {Y.}~\bibnamefont {Kobayashi}}, \ and\ \bibinfo {author}
  {\bibfnamefont {M.}~\bibnamefont {Sato}},\ }\href@noop {} {\bibfield
  {journal} {\bibinfo  {journal} {J. Phys. Soc. Jpn.},\ }\textbf {\bibinfo
  {volume} {76}},\ \bibinfo {pages} {023708} (\bibinfo {year}
  {2007}), cond-mat/0611659}\BibitemShut {NoStop}%
\bibitem [{\citenamefont {Seki}\ \emph {et~al.}(2010)\citenamefont {Seki},
  \citenamefont {Kurumaji}, \citenamefont {Ishiwata}, \citenamefont {Matsui},
  \citenamefont {Murakawa}, \citenamefont {Tokunaga}, \citenamefont {Kaneko},
  \citenamefont {Hasegawa},\ and\ \citenamefont {Tokura}}]{seki2010}%
  \BibitemOpen
  \bibfield  {author} {\bibinfo {author} {\bibfnamefont {S.}~\bibnamefont
  {Seki}}, \bibinfo {author} {\bibfnamefont {T.}~\bibnamefont {Kurumaji}},
  \bibinfo {author} {\bibfnamefont {S.}~\bibnamefont {Ishiwata}}, \bibinfo
  {author} {\bibfnamefont {H.}~\bibnamefont {Matsui}}, \bibinfo {author}
  {\bibfnamefont {H.}~\bibnamefont {Murakawa}}, \bibinfo {author}
  {\bibfnamefont {Y.}~\bibnamefont {Tokunaga}}, \bibinfo {author}
  {\bibfnamefont {Y.}~\bibnamefont {Kaneko}}, \bibinfo {author} {\bibfnamefont
  {T.}~\bibnamefont {Hasegawa}}, \ and\ \bibinfo {author} {\bibfnamefont
  {Y.}~\bibnamefont {Tokura}},\ }\href@noop {} {\bibfield  {journal} {\bibinfo
  {journal} {Phys. Rev. B},\ }\textbf {\bibinfo {volume} {82}},\ \bibinfo
  {pages} {064424} (\bibinfo {year} {2010}), arXiv:1008.5226 }\BibitemShut {NoStop}%
\bibitem [{\citenamefont {Moskvin}\ \emph {et~al.}(2009)\citenamefont
  {Moskvin}, \citenamefont {Panov},\ and\ \citenamefont {Drechsler}}]{moskvin}%
  \BibitemOpen
  \bibfield  {author} {\bibinfo {author} {\bibfnamefont {A.~S.}\ \bibnamefont
  {Moskvin}}, \bibinfo {author} {\bibfnamefont {Y.~D.}\ \bibnamefont {Panov}},
  \ and\ \bibinfo {author} {\bibfnamefont {S.-L.}\ \bibnamefont {Drechsler}},\
  }\href@noop {} {\bibfield  {journal} {\bibinfo  {journal} {Phys. Rev. B},\
  }\textbf {\bibinfo {volume} {79}},\ \bibinfo {pages} {104112} (\bibinfo
  {year} {2009}), arXiv:0801.1975}\BibitemShut {NoStop}%
\bibitem [{\citenamefont {Moskvin}\ and\ \citenamefont
  {Drechsler}(2008)}]{moskvin-2}%
  \BibitemOpen
  \bibfield  {author} {\bibinfo {author} {\bibfnamefont {A.~S.}\ \bibnamefont
  {Moskvin}}\ and\ \bibinfo {author} {\bibfnamefont {S.-L.}\ \bibnamefont
  {Drechsler}},\ }\href@noop {} {\bibfield  {journal} {\bibinfo  {journal}
  {Europhys. Lett.},\ }\textbf {\bibinfo {volume} {81}},\ \bibinfo {pages}
  {57004} (\bibinfo {year} {2008}), arXiv:0801.1102}\BibitemShut {NoStop}%
\bibitem [{\citenamefont {Affleck}\ \emph {et~al.}(1994)\citenamefont
  {Affleck}, \citenamefont {Gelfand},\ and\ \citenamefont
  {Singh}}]{affleck1994}%
  \BibitemOpen
  \bibfield  {author} {\bibinfo {author} {\bibfnamefont {I.}~\bibnamefont
  {Affleck}}, \bibinfo {author} {\bibfnamefont {M.~P.}\ \bibnamefont
  {Gelfand}}, \ and\ \bibinfo {author} {\bibfnamefont {R.~R.~P.}\ \bibnamefont
  {Singh}},\ }\href@noop {} {\bibfield  {journal} {\bibinfo  {journal} {J.
  Phys. A},\ }\textbf {\bibinfo {volume} {27}},\ \bibinfo {pages} {7313}
  (\bibinfo {year} {1994}), cond-mat/9408062}\BibitemShut {NoStop}%
\bibitem [{\citenamefont {Sandvik}(1999)}]{sandvik1999}%
  \BibitemOpen
  \bibfield  {author} {\bibinfo {author} {\bibfnamefont {A.~W.}\ \bibnamefont
  {Sandvik}},\ }\href@noop {} {\bibfield  {journal} {\bibinfo  {journal} {Phys.
  Rev. Lett.},\ }\textbf {\bibinfo {volume} {83}},\ \bibinfo {pages} {3069}
  (\bibinfo {year} {1999}), cond-mat/9904218}\BibitemShut {NoStop}%
\bibitem [{\citenamefont {Laflorencie}\ and\ \citenamefont
  {Poilblanc}(2003)}]{laflorencie2003}%
  \BibitemOpen
  \bibfield  {author} {\bibinfo {author} {\bibfnamefont {N.}~\bibnamefont
  {Laflorencie}}\ and\ \bibinfo {author} {\bibfnamefont {D.}~\bibnamefont
  {Poilblanc}},\ }\href@noop {} {\bibfield  {journal} {\bibinfo  {journal}
  {Phys. Rev. Lett.},\ }\textbf {\bibinfo {volume} {90}},\ \bibinfo {pages}
  {157202} (\bibinfo {year} {2003}), cond-mat/0211634}\BibitemShut {NoStop}%
\bibitem [{\citenamefont {Ueda}\ and\ \citenamefont
  {Totsuka}(2009)}]{ueda2009}%
  \BibitemOpen
  \bibfield  {author} {\bibinfo {author} {\bibfnamefont {H.~T.}\ \bibnamefont
  {Ueda}}\ and\ \bibinfo {author} {\bibfnamefont {K.}~\bibnamefont {Totsuka}},\
  }\href@noop {} {\bibfield  {journal} {\bibinfo  {journal} {Phys. Rev. B},\
  }\textbf {\bibinfo {volume} {80}},\ \bibinfo {pages} {014417} (\bibinfo
  {year} {2009}), arXiv:0905.0249}\BibitemShut {NoStop}%
\bibitem [{\citenamefont {Zhitomirsky}\ and\ \citenamefont
  {Tsunetsugu}(2010)}]{zhitomirsky}%
  \BibitemOpen
  \bibfield  {author} {\bibinfo {author} {\bibfnamefont {M.~E.}\ \bibnamefont
  {Zhitomirsky}}\ and\ \bibinfo {author} {\bibfnamefont {H.}~\bibnamefont
  {Tsunetsugu}},\ }\href@noop {} {\bibfield  {journal} {\bibinfo  {journal}
  {Europhys. Lett.},\ }\textbf {\bibinfo {volume} {92}},\ \bibinfo {pages}
  {37001} (\bibinfo {year} {2010}), arXiv:1003.4096 }\BibitemShut {NoStop}%
\bibitem [{\citenamefont {Nishimoto}\ \emph {et~al.}(2010)\citenamefont
  {Nishimoto}, \citenamefont {Drechsler}, \citenamefont {Kuzian}, \citenamefont
  {{van der Brink}},\ and\ \citenamefont {Richter}}]{nishimoto2010}%
  \BibitemOpen
  \bibfield  {author} {\bibinfo {author} {\bibfnamefont {S.}~\bibnamefont
  {Nishimoto}}, \bibinfo {author} {\bibfnamefont {S.-L.}\ \bibnamefont
  {Drechsler}}, \bibinfo {author} {\bibfnamefont {R.}~\bibnamefont {Kuzian}},
  \bibinfo {author} {\bibfnamefont {J.}~\bibnamefont {{van der Brink}}}, \ and\
  \bibinfo {author} {\bibfnamefont {J.}~\bibnamefont {Richter}},\ }\href@noop
  {} {} (\bibinfo {year} {2010}),\ \Eprint {http://arxiv.org/abs/1005.5500}
  {arXiv:1005.5500} \BibitemShut {NoStop}%
\bibitem [{\citenamefont {Zinke}\ \emph {et~al.}(2009)\citenamefont {Zinke},
  \citenamefont {Drechsler},\ and\ \citenamefont {Richter}}]{zinke2009}%
  \BibitemOpen
  \bibfield  {author} {\bibinfo {author} {\bibfnamefont {R.}~\bibnamefont
  {Zinke}}, \bibinfo {author} {\bibfnamefont {S.-L.}\ \bibnamefont
  {Drechsler}}, \ and\ \bibinfo {author} {\bibfnamefont {J.}~\bibnamefont
  {Richter}},\ }\href@noop {} {\bibfield  {journal} {\bibinfo  {journal} {Phys.
  Rev. B},\ }\textbf {\bibinfo {volume} {79}},\ \bibinfo {pages} {094425}
  (\bibinfo {year} {2009}), arXiv:0807.3431}\BibitemShut {NoStop}%
\bibitem [{\citenamefont {Hegenbart}\ \emph {et~al.}(1981)\citenamefont
  {Hegenbart}, \citenamefont {Rau},\ and\ \citenamefont {Range}}]{cu2geo4-str}%
  \BibitemOpen
  \bibfield  {author} {\bibinfo {author} {\bibfnamefont {W.}~\bibnamefont
  {Hegenbart}}, \bibinfo {author} {\bibfnamefont {F.}~\bibnamefont {Rau}}, \
  and\ \bibinfo {author} {\bibfnamefont {K.-J.}\ \bibnamefont {Range}},\
  }\href@noop {} {\bibfield  {journal} {\bibinfo  {journal} {Mater. Res.
  Bull.},\ }\textbf {\bibinfo {volume} {16}},\ \bibinfo {pages} {413} (\bibinfo
  {year} {1981})}\BibitemShut {NoStop}%
\bibitem [{\citenamefont {Yamada}\ \emph {et~al.}(2000)\citenamefont {Yamada},
  \citenamefont {Hiroi}, \citenamefont {Takano}, \citenamefont {Nohara},\ and\
  \citenamefont {Takagi}}]{yamada2000}%
  \BibitemOpen
  \bibfield  {author} {\bibinfo {author} {\bibfnamefont {T.}~\bibnamefont
  {Yamada}}, \bibinfo {author} {\bibfnamefont {Z.}~\bibnamefont {Hiroi}},
  \bibinfo {author} {\bibfnamefont {M.}~\bibnamefont {Takano}}, \bibinfo
  {author} {\bibfnamefont {M.}~\bibnamefont {Nohara}}, \ and\ \bibinfo {author}
  {\bibfnamefont {H.}~\bibnamefont {Takagi}},\ }\href@noop {} {\bibfield
  {journal} {\bibinfo  {journal} {J. Phys. Soc. Jpn.},\ }\textbf {\bibinfo
  {volume} {69}},\ \bibinfo {pages} {1477} (\bibinfo {year}
  {2000})}\BibitemShut {NoStop}%
\bibitem [{\citenamefont {Starykh}\ \emph {et~al.}(2005)\citenamefont
  {Starykh}, \citenamefont {Furusaki},\ and\ \citenamefont
  {Balents}}]{starykh2005}%
  \BibitemOpen
  \bibfield  {author} {\bibinfo {author} {\bibfnamefont {O.~A.}\ \bibnamefont
  {Starykh}}, \bibinfo {author} {\bibfnamefont {A.}~\bibnamefont {Furusaki}}, \
  and\ \bibinfo {author} {\bibfnamefont {L.}~\bibnamefont {Balents}},\
  }\href@noop {} {\bibfield  {journal} {\bibinfo  {journal} {Phys. Rev. B},\
  }\textbf {\bibinfo {volume} {72}},\ \bibinfo {pages} {094416} (\bibinfo
  {year} {2005}), cond-mat/0503296}\BibitemShut {NoStop}%
\bibitem [{\citenamefont {Koepernik}\ and\ \citenamefont
  {Eschrig}(1999)}]{fplo}%
  \BibitemOpen
  \bibfield  {author} {\bibinfo {author} {\bibfnamefont {K.}~\bibnamefont
  {Koepernik}}\ and\ \bibinfo {author} {\bibfnamefont {H.}~\bibnamefont
  {Eschrig}},\ }\href@noop {} {\bibfield  {journal} {\bibinfo  {journal} {Phys.
  Rev. B},\ }\textbf {\bibinfo {volume} {59}},\ \bibinfo {pages} {1743}
  (\bibinfo {year} {1999})}\BibitemShut {NoStop}%
\bibitem [{\citenamefont {Perdew}\ and\ \citenamefont {Wang}(1992)}]{pw92}%
  \BibitemOpen
  \bibfield  {author} {\bibinfo {author} {\bibfnamefont {J.~P.}\ \bibnamefont
  {Perdew}}\ and\ \bibinfo {author} {\bibfnamefont {Y.}~\bibnamefont {Wang}},\
  }\href@noop {} {\bibfield  {journal} {\bibinfo  {journal} {Phys. Rev. B},\
  }\textbf {\bibinfo {volume} {45}},\ \bibinfo {pages} {13244} (\bibinfo {year}
  {1992})}\BibitemShut {NoStop}%
\bibitem [{\citenamefont {Janson}\ \emph
  {et~al.}(2010){\natexlab{a}}\citenamefont {Janson}, \citenamefont {Tsirlin},
  \citenamefont {Schmitt},\ and\ \citenamefont {Rosner}}]{dioptase}%
  \BibitemOpen
  \bibfield  {author} {\bibinfo {author} {\bibfnamefont {O.}~\bibnamefont
  {Janson}}, \bibinfo {author} {\bibfnamefont {A.~A.}\ \bibnamefont {Tsirlin}},
  \bibinfo {author} {\bibfnamefont {M.}~\bibnamefont {Schmitt}}, \ and\
  \bibinfo {author} {\bibfnamefont {H.}~\bibnamefont {Rosner}},\ }\href@noop {}
  {\bibfield  {journal} {\bibinfo  {journal} {Phys. Rev. B},\ }\textbf
  {\bibinfo {volume} {82}},\ \bibinfo {pages} {014424} (\bibinfo {year}
  {2010}{\natexlab{a}}), arXiv:1004.3765}\BibitemShut {NoStop}%
\bibitem [{\citenamefont {Tsirlin}\ \emph {et~al.}(2010)\citenamefont
  {Tsirlin}, \citenamefont {Janson},\ and\ \citenamefont {Rosner}}]{cu2v2o7}%
  \BibitemOpen
  \bibfield  {author} {\bibinfo {author} {\bibfnamefont {A.~A.}\ \bibnamefont
  {Tsirlin}}, \bibinfo {author} {\bibfnamefont {O.}~\bibnamefont {Janson}}, \
  and\ \bibinfo {author} {\bibfnamefont {H.}~\bibnamefont {Rosner}},\
  }\href@noop {} {\bibfield  {journal} {\bibinfo  {journal} {Phys. Rev. B},\
  }\textbf {\bibinfo {volume} {82}},\ \bibinfo {pages} {144416} (\bibinfo
  {year} {2010}), arXiv:1007.1646}\BibitemShut {NoStop}%
\bibitem [{\citenamefont {Janson}\ \emph {et~al.}(2009)\citenamefont {Janson},
  \citenamefont {Schnelle}, \citenamefont {Schmidt}, \citenamefont {Prots},
  \citenamefont {Drechsler}, \citenamefont {Filatov},\ and\ \citenamefont
  {Rosner}}]{janson2009}%
  \BibitemOpen
  \bibfield  {author} {\bibinfo {author} {\bibfnamefont {O.}~\bibnamefont
  {Janson}}, \bibinfo {author} {\bibfnamefont {W.}~\bibnamefont {Schnelle}},
  \bibinfo {author} {\bibfnamefont {M.}~\bibnamefont {Schmidt}}, \bibinfo
  {author} {\bibfnamefont {Y.}~\bibnamefont {Prots}}, \bibinfo {author}
  {\bibfnamefont {S.-L.}\ \bibnamefont {Drechsler}}, \bibinfo {author}
  {\bibfnamefont {S.~K.}\ \bibnamefont {Filatov}}, \ and\ \bibinfo {author}
  {\bibfnamefont {H.}~\bibnamefont {Rosner}},\ }\href@noop {} {\bibfield
  {journal} {\bibinfo  {journal} {New J. Phys.},\ }\textbf {\bibinfo {volume}
  {11}},\ \bibinfo {pages} {113034} (\bibinfo {year} {2009}), arXiv:0907.4874}\BibitemShut
  {NoStop}%
\bibitem [{\citenamefont {Janson}\ \emph
  {et~al.}(2010){\natexlab{b}}\citenamefont {Janson}, \citenamefont {Richter},
  \citenamefont {Sindzingre},\ and\ \citenamefont {Rosner}}]{volborthite}%
  \BibitemOpen
  \bibfield  {author} {\bibinfo {author} {\bibfnamefont {O.}~\bibnamefont
  {Janson}}, \bibinfo {author} {\bibfnamefont {J.}~\bibnamefont {Richter}},
  \bibinfo {author} {\bibfnamefont {P.}~\bibnamefont {Sindzingre}}, \ and\
  \bibinfo {author} {\bibfnamefont {H.}~\bibnamefont {Rosner}},\ }\href@noop {}
  {\bibfield  {journal} {\bibinfo  {journal} {Phys. Rev. B},\ }\textbf
  {\bibinfo {volume} {82}},\ \bibinfo {pages} {104434} (\bibinfo {year}
  {2010}{\natexlab{b}}), arXiv:1004.2185}\BibitemShut {NoStop}%
\bibitem [{\citenamefont {Eschrig}\ and\ \citenamefont
  {Koepernik}(2009)}]{wannier}%
  \BibitemOpen
  \bibfield  {author} {\bibinfo {author} {\bibfnamefont {H.}~\bibnamefont
  {Eschrig}}\ and\ \bibinfo {author} {\bibfnamefont {K.}~\bibnamefont
  {Koepernik}},\ }\href@noop {} {\bibfield  {journal} {\bibinfo  {journal}
  {Phys. Rev. B},\ }\textbf {\bibinfo {volume} {80}},\ \bibinfo {pages}
  {104503} (\bibinfo {year} {2009}), arXiv:0905.4844}\BibitemShut {NoStop}%
\bibitem [{not(){\natexlab{a}}}]{note1}%
  \BibitemOpen
  \bibinfo {note} {The symmetry of the model restricts the computationally
  feasible size of the finite lattice to $4\times 4$, with two $J_2$ bonds
  only.}\BibitemShut {Stop}%
\bibitem [{\citenamefont {Johnston}\ \emph {et~al.}(2000)\citenamefont
  {Johnston}, \citenamefont {Kremer}, \citenamefont {Troyer}, \citenamefont
  {Wang}, \citenamefont {Kl\"umper}, \citenamefont {Bud'ko}, \citenamefont
  {Panchula},\ and\ \citenamefont {Canfield}}]{johnston2000}%
  \BibitemOpen
  \bibfield  {author} {\bibinfo {author} {\bibfnamefont {D.~C.}\ \bibnamefont
  {Johnston}}, \bibinfo {author} {\bibfnamefont {R.~K.}\ \bibnamefont
  {Kremer}}, \bibinfo {author} {\bibfnamefont {M.}~\bibnamefont {Troyer}},
  \bibinfo {author} {\bibfnamefont {X.}~\bibnamefont {Wang}}, \bibinfo {author}
  {\bibfnamefont {A.}~\bibnamefont {Kl\"umper}}, \bibinfo {author}
  {\bibfnamefont {S.~L.}\ \bibnamefont {Bud'ko}}, \bibinfo {author}
  {\bibfnamefont {A.~F.}\ \bibnamefont {Panchula}}, \ and\ \bibinfo {author}
  {\bibfnamefont {P.~C.}\ \bibnamefont {Canfield}},\ }\href@noop {} {\bibfield
  {journal} {\bibinfo  {journal} {Phys. Rev. B},\ }\textbf {\bibinfo {volume}
  {61}},\ \bibinfo {pages} {9558} (\bibinfo {year} {2000}), cond-mat/0003271}\BibitemShut
  {NoStop}%
\bibitem [{\citenamefont {Garrett}\ \emph {et~al.}(1997)\citenamefont
  {Garrett}, \citenamefont {Nagler}, \citenamefont {Tennant}, \citenamefont
  {Sales},\ and\ \citenamefont {Barnes}}]{garrett1997}%
  \BibitemOpen
  \bibfield  {author} {\bibinfo {author} {\bibfnamefont {A.~W.}\ \bibnamefont
  {Garrett}}, \bibinfo {author} {\bibfnamefont {S.~E.}\ \bibnamefont {Nagler}},
  \bibinfo {author} {\bibfnamefont {D.~A.}\ \bibnamefont {Tennant}}, \bibinfo
  {author} {\bibfnamefont {B.~C.}\ \bibnamefont {Sales}}, \ and\ \bibinfo
  {author} {\bibfnamefont {T.}~\bibnamefont {Barnes}},\ }\href@noop {}
  {\bibfield  {journal} {\bibinfo  {journal} {Phys. Rev. Lett.},\ }\textbf
  {\bibinfo {volume} {79}},\ \bibinfo {pages} {745} (\bibinfo {year}
  {1997}), cond-mat/9704092}\BibitemShut {NoStop}%
\bibitem [{\citenamefont {Kaul}\ \emph {et~al.}(2003)\citenamefont {Kaul},
  \citenamefont {Rosner}, \citenamefont {Yushankhai}, \citenamefont
  {Sichelschmidt}, \citenamefont {Shpanchenko},\ and\ \citenamefont
  {Geibel}}]{kaul2003}%
  \BibitemOpen
  \bibfield  {author} {\bibinfo {author} {\bibfnamefont {E.~E.}\ \bibnamefont
  {Kaul}}, \bibinfo {author} {\bibfnamefont {H.}~\bibnamefont {Rosner}},
  \bibinfo {author} {\bibfnamefont {V.}~\bibnamefont {Yushankhai}}, \bibinfo
  {author} {\bibfnamefont {J.}~\bibnamefont {Sichelschmidt}}, \bibinfo {author}
  {\bibfnamefont {R.~V.}\ \bibnamefont {Shpanchenko}}, \ and\ \bibinfo {author}
  {\bibfnamefont {C.}~\bibnamefont {Geibel}},\ }\href@noop {} {\bibfield
  {journal} {\bibinfo  {journal} {Phys. Rev. B},\ }\textbf {\bibinfo {volume}
  {67}},\ \bibinfo {pages} {174417} (\bibinfo {year} {2003}), cond-mat/0209409}\BibitemShut
  {NoStop}%
\bibitem [{\citenamefont {Tsirlin}\ and\ \citenamefont
  {Rosner}(2011)}]{pb2v3o9}%
  \BibitemOpen
  \bibfield  {author} {\bibinfo {author} {\bibfnamefont {A.~A.}\ \bibnamefont
  {Tsirlin}}\ and\ \bibinfo {author} {\bibfnamefont {H.}~\bibnamefont
  {Rosner}},\ }\href@noop {} {\bibfield  {journal} {\bibinfo  {journal} {Phys.
  Rev. B},\ }\textbf {\bibinfo {volume} {83}},\ \bibinfo {pages} {064415}
  (\bibinfo {year} {2011}), arXiv:1011.3981}\BibitemShut {NoStop}%
\bibitem [{\citenamefont {Bursill}\ \emph {et~al.}(1995)\citenamefont
  {Bursill}, \citenamefont {Gehring}, \citenamefont {Farnell}, \citenamefont
  {Parkinson}, \citenamefont {Xiang},\ and\ \citenamefont {Zeng}}]{bursill}%
  \BibitemOpen
  \bibfield  {author} {\bibinfo {author} {\bibfnamefont {R.}~\bibnamefont
  {Bursill}}, \bibinfo {author} {\bibfnamefont {G.~A.}\ \bibnamefont
  {Gehring}}, \bibinfo {author} {\bibfnamefont {D.~J.~J.}\ \bibnamefont
  {Farnell}}, \bibinfo {author} {\bibfnamefont {J.~B.}\ \bibnamefont
  {Parkinson}}, \bibinfo {author} {\bibfnamefont {T.}~\bibnamefont {Xiang}}, \
  and\ \bibinfo {author} {\bibfnamefont {C.}~\bibnamefont {Zeng}},\ }\href@noop
  {} {\bibfield  {journal} {\bibinfo  {journal} {J. Phys.: Condens. Matter},\
  }\textbf {\bibinfo {volume} {7}},\ \bibinfo {pages} {8605} (\bibinfo {year}
  {1995}), cond-mat/9511044}\BibitemShut {NoStop}%
\bibitem [{\citenamefont {Bishop}\ \emph {et~al.}(1998)\citenamefont {Bishop},
  \citenamefont {Farnell},\ and\ \citenamefont {Parkinson}}]{bishop98}%
  \BibitemOpen
  \bibfield  {author} {\bibinfo {author} {\bibfnamefont {R.~F.}\ \bibnamefont
  {Bishop}}, \bibinfo {author} {\bibfnamefont {D.~J.~J.}\ \bibnamefont
  {Farnell}}, \ and\ \bibinfo {author} {\bibfnamefont {J.~B.}\ \bibnamefont
  {Parkinson}},\ }\href@noop {} {\bibfield  {journal} {\bibinfo  {journal}
  {Phys. Rev. B},\ }\textbf {\bibinfo {volume} {58}},\ \bibinfo {pages} {6394}
  (\bibinfo {year} {1998}), cond-mat/9804079}\BibitemShut {NoStop}%
\bibitem [{\citenamefont {Kr\"uger}\ \emph {et~al.}(2000)\citenamefont
  {Kr\"uger}, \citenamefont {Richter}, \citenamefont {Schulenburg},
  \citenamefont {Farnell},\ and\ \citenamefont {Bishop}}]{krueger00}%
  \BibitemOpen
  \bibfield  {author} {\bibinfo {author} {\bibfnamefont {S.~E.}\ \bibnamefont
  {Kr\"uger}}, \bibinfo {author} {\bibfnamefont {J.}~\bibnamefont {Richter}},
  \bibinfo {author} {\bibfnamefont {J.}~\bibnamefont {Schulenburg}}, \bibinfo
  {author} {\bibfnamefont {D.~J.~J.}\ \bibnamefont {Farnell}}, \ and\ \bibinfo
  {author} {\bibfnamefont {R.~F.}\ \bibnamefont {Bishop}},\ }\href@noop {}
  {\bibfield  {journal} {\bibinfo  {journal} {Phys. Rev. B},\ }\textbf
  {\bibinfo {volume} {61}},\ \bibinfo {pages} {14607} (\bibinfo {year}
  {2000}), cond-mat/0003126}\BibitemShut {NoStop}%
\bibitem [{\citenamefont {Kr\"uger}\ and\ \citenamefont
  {Richter}(2001)}]{krueger01}%
  \BibitemOpen
  \bibfield  {author} {\bibinfo {author} {\bibfnamefont {S.~E.}\ \bibnamefont
  {Kr\"uger}}\ and\ \bibinfo {author} {\bibfnamefont {J.}~\bibnamefont
  {Richter}},\ }\href@noop {} {\bibfield  {journal} {\bibinfo  {journal} {Phys.
  Rev. B},\ }\textbf {\bibinfo {volume} {64}},\ \bibinfo {pages} {024433}
  (\bibinfo {year} {2001}), cond-mat/0104540}\BibitemShut {NoStop}%
\bibitem [{\citenamefont {Farnell}\ and\ \citenamefont
  {Bishop}(2004)}]{farnell04}%
  \BibitemOpen
  \bibfield  {author} {\bibinfo {author} {\bibfnamefont {D.~J.~J.}\
  \bibnamefont {Farnell}}\ and\ \bibinfo {author} {\bibfnamefont {R.~F.}\
  \bibnamefont {Bishop}},\ }\href@noop {} {\bibfield  {journal} {\bibinfo
  {journal} {Lecture Notes in Physics},\ }\textbf {\bibinfo {volume} {645}},\
  \bibinfo {pages} {307} (\bibinfo {year} {2004})}\BibitemShut {NoStop}%
\bibitem [{\citenamefont {Darradi}\ \emph {et~al.}(2005)\citenamefont
  {Darradi}, \citenamefont {Richter},\ and\ \citenamefont
  {Farnell}}]{darradi_shastry}%
  \BibitemOpen
  \bibfield  {author} {\bibinfo {author} {\bibfnamefont {R.}~\bibnamefont
  {Darradi}}, \bibinfo {author} {\bibfnamefont {J.}~\bibnamefont {Richter}}, \
  and\ \bibinfo {author} {\bibfnamefont {D.~J.~J.}\ \bibnamefont {Farnell}},\
  }\href@noop {} {\bibfield  {journal} {\bibinfo  {journal} {Phys. Rev. B},\
  }\textbf {\bibinfo {volume} {72}},\ \bibinfo {pages} {104425} (\bibinfo
  {year} {2005}), cond-mat/0504283}\BibitemShut {NoStop}%
\bibitem [{\citenamefont {Schmalfu\ss}\ \emph {et~al.}(2006)\citenamefont
  {Schmalfu\ss}, \citenamefont {Darradi}, \citenamefont {Richter},
  \citenamefont {Schulenburg},\ and\ \citenamefont {Ihle}}]{Schm:2006}%
  \BibitemOpen
  \bibfield  {author} {\bibinfo {author} {\bibfnamefont {D.}~\bibnamefont
  {Schmalfu\ss}}, \bibinfo {author} {\bibfnamefont {R.}~\bibnamefont
  {Darradi}}, \bibinfo {author} {\bibfnamefont {J.}~\bibnamefont {Richter}},
  \bibinfo {author} {\bibfnamefont {J.}~\bibnamefont {Schulenburg}}, \ and\
  \bibinfo {author} {\bibfnamefont {D.}~\bibnamefont {Ihle}},\ }\href@noop {}
  {\bibfield  {journal} {\bibinfo  {journal} {Phys. Rev. Lett.},\ }\textbf
  {\bibinfo {volume} {97}},\ \bibinfo {pages} {157201} (\bibinfo {year}
  {2006}), cond-mat/0604172}\BibitemShut {NoStop}%
\bibitem [{\citenamefont {Darradi}\ \emph {et~al.}(2008)\citenamefont
  {Darradi}, \citenamefont {Derzhko}, \citenamefont {Zinke}, \citenamefont
  {Schulenburg}, \citenamefont {Kr\"uger},\ and\ \citenamefont
  {Richter}}]{darradi08}%
  \BibitemOpen
  \bibfield  {author} {\bibinfo {author} {\bibfnamefont {R.}~\bibnamefont
  {Darradi}}, \bibinfo {author} {\bibfnamefont {O.}~\bibnamefont {Derzhko}},
  \bibinfo {author} {\bibfnamefont {R.}~\bibnamefont {Zinke}}, \bibinfo
  {author} {\bibfnamefont {J.}~\bibnamefont {Schulenburg}}, \bibinfo {author}
  {\bibfnamefont {S.~E.}\ \bibnamefont {Kr\"uger}}, \ and\ \bibinfo {author}
  {\bibfnamefont {J.}~\bibnamefont {Richter}},\ }\href@noop {} {\bibfield
  {journal} {\bibinfo  {journal} {Phys. Rev. B},\ }\textbf {\bibinfo {volume}
  {78}},\ \bibinfo {pages} {214415} (\bibinfo {year} {2008}), arXiv:0806.3825}\BibitemShut
  {NoStop}%
\bibitem [{\citenamefont {Bishop}\ \emph {et~al.}(2008)\citenamefont {Bishop},
  \citenamefont {Li}, \citenamefont {Darradi},\ and\ \citenamefont
  {Richter}}]{bishop08}%
  \BibitemOpen
  \bibfield  {author} {\bibinfo {author} {\bibfnamefont {R.~F.}\ \bibnamefont
  {Bishop}}, \bibinfo {author} {\bibfnamefont {P.~H.~Y.}\ \bibnamefont {Li}},
  \bibinfo {author} {\bibfnamefont {R.}~\bibnamefont {Darradi}}, \ and\
  \bibinfo {author} {\bibfnamefont {J.}~\bibnamefont {Richter}},\ }\href@noop
  {} {\bibfield  {journal} {\bibinfo  {journal} {J. Phys.: Cond. Matter},\
  }\textbf {\bibinfo {volume} {20}},\ \bibinfo {pages} {255251} (\bibinfo
  {year} {2008}), arXiv:0705.2201}\BibitemShut {NoStop}%
\bibitem [{\citenamefont {Bishop}\ \emph {et~al.}(2009)\citenamefont {Bishop},
  \citenamefont {Li}, \citenamefont {Farnell},\ and\ \citenamefont
  {Campbell}}]{bishop09}%
  \BibitemOpen
  \bibfield  {author} {\bibinfo {author} {\bibfnamefont {R.~F.}\ \bibnamefont
  {Bishop}}, \bibinfo {author} {\bibfnamefont {P.~H.~Y.}\ \bibnamefont {Li}},
  \bibinfo {author} {\bibfnamefont {D.~J.~J.}\ \bibnamefont {Farnell}}, \ and\
  \bibinfo {author} {\bibfnamefont {C.~E.}\ \bibnamefont {Campbell}},\
  }\href@noop {} {\bibfield  {journal} {\bibinfo  {journal} {Phys. Rev. B},\
  }\textbf {\bibinfo {volume} {79}},\ \bibinfo {pages} {174405} (\bibinfo
  {year} {2009}), arXiv:0812.3821}\BibitemShut {NoStop}%
\bibitem [{\citenamefont {Richter}\ \emph {et~al.}(2010)\citenamefont
  {Richter}, \citenamefont {Darradi}, \citenamefont {Schulenburg},
  \citenamefont {Farnell},\ and\ \citenamefont {Rosner}}]{richter10}%
  \BibitemOpen
  \bibfield  {author} {\bibinfo {author} {\bibfnamefont {J.}~\bibnamefont
  {Richter}}, \bibinfo {author} {\bibfnamefont {R.}~\bibnamefont {Darradi}},
  \bibinfo {author} {\bibfnamefont {J.}~\bibnamefont {Schulenburg}}, \bibinfo
  {author} {\bibfnamefont {D.~J.~J.}\ \bibnamefont {Farnell}}, \ and\ \bibinfo
  {author} {\bibfnamefont {H.}~\bibnamefont {Rosner}},\ }\href@noop {}
  {\bibfield  {journal} {\bibinfo  {journal} {Phys. Rev. B},\ }\textbf
  {\bibinfo {volume} {81}},\ \bibinfo {pages} {174429} (\bibinfo {year}
  {2010}), arXiv:1002.2299}\BibitemShut {NoStop}%
\bibitem [{\citenamefont {Zinke}\ \emph {et~al.}(2010)\citenamefont {Zinke},
  \citenamefont {Richter},\ and\ \citenamefont {Drechsler}}]{zinke10}%
  \BibitemOpen
  \bibfield  {author} {\bibinfo {author} {\bibfnamefont {R.}~\bibnamefont
  {Zinke}}, \bibinfo {author} {\bibfnamefont {J.}~\bibnamefont {Richter}}, \
  and\ \bibinfo {author} {\bibfnamefont {S.-L.}\ \bibnamefont {Drechsler}},\
  }\href@noop {} {\bibfield  {journal} {\bibinfo  {journal} {J. Phys.: Condens.
  Matter},\ }\textbf {\bibinfo {volume} {22}},\ \bibinfo {pages} {446002}
  (\bibinfo {year} {2010}), arXiv:1008.0317}\BibitemShut {NoStop}%
\bibitem [{\citenamefont {Bishop}\ \emph {et~al.}(2000)\citenamefont {Bishop},
  \citenamefont {Farnell}, \citenamefont {Kr\"uger}, \citenamefont {Parkinson},
  \citenamefont {Richter},\ and\ \citenamefont {Zeng}}]{bishop00}%
  \BibitemOpen
  \bibfield  {author} {\bibinfo {author} {\bibfnamefont {R.~F.}\ \bibnamefont
  {Bishop}}, \bibinfo {author} {\bibfnamefont {D.~J.~J.}\ \bibnamefont
  {Farnell}}, \bibinfo {author} {\bibfnamefont {S.~E.}\ \bibnamefont
  {Kr\"uger}}, \bibinfo {author} {\bibfnamefont {J.~B.}\ \bibnamefont
  {Parkinson}}, \bibinfo {author} {\bibfnamefont {J.}~\bibnamefont {Richter}},
  \ and\ \bibinfo {author} {\bibfnamefont {C.}~\bibnamefont {Zeng}},\
  }\href@noop {} {\bibfield  {journal} {\bibinfo  {journal} {J. Phys.: Condens.
  Matter},\ }\textbf {\bibinfo {volume} {12}},\ \bibinfo {pages} {6887}
  (\bibinfo {year} {2000}), cond-mat/0011008}\BibitemShut {NoStop}%
\bibitem [{ccc()}]{cccm}%
  \BibitemOpen
  \bibinfo {note} {See
  http://www-e.uni-magdeburg.de/jschulen/ccm/index.html}\BibitemShut {NoStop}%
\bibitem [{not(){\natexlab{b}}}]{note2}%
  \BibitemOpen
  \bibinfo {note} {The $b'$ parameter of the spin lattice is twice smaller than
  the crystallographic lattice parameter in the $ab$ plane.}\BibitemShut
  {Stop}%
\bibitem [{\citenamefont {Valenti}\ \emph {et~al.}(2003)\citenamefont
  {Valenti}, \citenamefont {Saha-Dasgupta}, \citenamefont {Gros},\ and\
  \citenamefont {Rosner}}]{valenti2003}%
  \BibitemOpen
  \bibfield  {author} {\bibinfo {author} {\bibfnamefont {R.}~\bibnamefont
  {Valenti}}, \bibinfo {author} {\bibfnamefont {T.}~\bibnamefont
  {Saha-Dasgupta}}, \bibinfo {author} {\bibfnamefont {C.}~\bibnamefont {Gros}},
  \ and\ \bibinfo {author} {\bibfnamefont {H.}~\bibnamefont {Rosner}},\
  }\href@noop {} {\bibfield  {journal} {\bibinfo  {journal} {Phys. Rev. B},\
  }\textbf {\bibinfo {volume} {67}},\ \bibinfo {pages} {245110} (\bibinfo
  {year} {2003}), cond-mat/0301119}\BibitemShut {NoStop}%
\bibitem [{\citenamefont {Tsirlin}\ and\ \citenamefont
  {Rosner}(2010){\natexlab{c}}}]{cucl}%
  \BibitemOpen
  \bibfield  {author} {\bibinfo {author} {\bibfnamefont {A.~A.}\ \bibnamefont
  {Tsirlin}}\ and\ \bibinfo {author} {\bibfnamefont {H.}~\bibnamefont
  {Rosner}},\ }\href@noop {} {\bibfield  {journal} {\bibinfo  {journal} {Phys.
  Rev. B},\ }\textbf {\bibinfo {volume} {82}},\ \bibinfo {pages} {060409(R)}
  (\bibinfo {year} {2010}{\natexlab{c}}), arXiv:1007.3883}\BibitemShut {NoStop}%
\bibitem [{\citenamefont {Sindzingre}\ \emph {et~al.}(2002)\citenamefont
  {Sindzingre}, \citenamefont {Fouet},\ and\ \citenamefont
  {Lhuillier}}]{sindzingre2002}%
  \BibitemOpen
  \bibfield  {author} {\bibinfo {author} {\bibfnamefont {P.}~\bibnamefont
  {Sindzingre}}, \bibinfo {author} {\bibfnamefont {J.-B.}\ \bibnamefont
  {Fouet}}, \ and\ \bibinfo {author} {\bibfnamefont {C.}~\bibnamefont
  {Lhuillier}},\ }\href@noop {} {\bibfield  {journal} {\bibinfo  {journal}
  {Phys. Rev. B},\ }\textbf {\bibinfo {volume} {66}},\ \bibinfo {pages}
  {174424} (\bibinfo {year} {2002}), cond-mat/0204299}\BibitemShut {NoStop}%
\bibitem [{\citenamefont {Yasuda}\ \emph {et~al.}(2005)\citenamefont {Yasuda},
  \citenamefont {Todo}, \citenamefont {Hukushima}, \citenamefont {Alet},
  \citenamefont {Keller}, \citenamefont {Troyer},\ and\ \citenamefont
  {Takayama}}]{yasuda2005}%
  \BibitemOpen
  \bibfield  {author} {\bibinfo {author} {\bibfnamefont {C.}~\bibnamefont
  {Yasuda}}, \bibinfo {author} {\bibfnamefont {S.}~\bibnamefont {Todo}},
  \bibinfo {author} {\bibfnamefont {K.}~\bibnamefont {Hukushima}}, \bibinfo
  {author} {\bibfnamefont {F.}~\bibnamefont {Alet}}, \bibinfo {author}
  {\bibfnamefont {M.}~\bibnamefont {Keller}}, \bibinfo {author} {\bibfnamefont
  {M.}~\bibnamefont {Troyer}}, \ and\ \bibinfo {author} {\bibfnamefont
  {H.}~\bibnamefont {Takayama}},\ }\href@noop {} {\bibfield  {journal}
  {\bibinfo  {journal} {Phys. Rev. Lett.},\ }\textbf {\bibinfo {volume} {94}},\
  \bibinfo {pages} {217201} (\bibinfo {year} {2005}), cond-mat/0312392}\BibitemShut {NoStop}%
\end{thebibliography}
%

\end{document}